\documentclass[12pt]{iopart}

\pdfoutput=1

\usepackage{graphicx,sidecap}
\usepackage{bm}
\usepackage{braket}
\usepackage{amssymb}
\usepackage{mathrsfs}

\expandafter\let\csname equation*\endcsname\relax
\expandafter\let\csname endequation*\endcsname\relax

\usepackage{amsmath}
\usepackage{xcolor}
\usepackage{hyperref}
\usepackage{enumerate}
\usepackage[toc,title]{appendix}
\usepackage{verbatim}
\usepackage[margin=30pt, bf, font=footnotesize, center, justification=justified]{caption}[2004/07/16]

\hypersetup{colorlinks,bookmarksopen,bookmarksnumbered,
citecolor=black,
linkcolor=black,
pdfstartview=green,
urlcolor=purple}

\catcode`@=11 
\renewcommand\tableofcontents{%
  \section*{\contentsname}%
  \@starttoc{toc}%
}
\catcode`@=12

\def\be{\begin{equation}}
\def\ee{\end{equation}}

\def\bea{\begin{eqnarray}}
\def\eea{\end{eqnarray}}

\begin{document}

\title[Entanglement hamiltonian evolution during thermalization  in CFT]
{Entanglement hamiltonian evolution during thermalization in conformal field theory}

\author{Xueda Wen$^{1}$, Shinsei Ryu$^{2}$ and Andreas W.W. Ludwig$^3$}
\address{$^1$\,Department of Physics, Massachusetts Institute of Technology, Cambridge, MA 02139, USA}
\address{$^2$\, James Franck Institute and Kadanoff Center for Theoretical Physics, University of Chicago, Illinois 60637, USA}
\address{$^3$\, Department of Physics, University of California, Santa Barbara, CA 93106, USA}

\vspace{10pt}

\vspace{.5cm}

\begin{abstract}
In this work, we study the time evolution of the entanglement hamiltonian during the process of thermalization in a
(1+1)-dimensional conformal field theory (CFT) after a quantum quench from a special class of initial states.
In particular, we focus on a 
subsystem which is 
 a finite interval at the end of a semi-infinite line.
Based on conformal mappings, the exact forms of
both entanglement hamiltonian and entanglement spectrum of the subsystem can be obtained.
Aside from
various interesting features, it is found that
 in the infinite time limit
 the entanglement hamiltonian and entanglement spectrum
are exactly the same as those in the thermal ensemble.
The entanglement spectrum approaches the   steady state  spectrum  exponentially in time.
We also study the modular flows generated by the entanglement hamiltonian in Minkowski spacetime,
which provides us with an intuitive picture
of  how the entanglement propagates and how the subsystem is thermalized.
Furthermore, the effect of a generic initial state is also discussed.
\end{abstract}

\maketitle

 \tableofcontents

\section{Introduction}

\subsection{Introduction}


\paragraph{Quantum quench and thermalization}
Unraveling non-equilibrium dynamics in quantum many-body systems remains an important open question.
A paradigmatic protocol for non-equilibrium dynamics, which will be our focus
here, is a  {\it quantum quench}, in which one changes  a system's hamiltonian as a function of time.
In typical settings,
the time-dependent hamiltonian $H(t)$ interpolates between  two hamiltonians,
$H(t \to -\infty)\equiv H_i$ and $H(t\to \infty)\equiv H_{f}$,
and the change in the hamiltonian occurs 
during  a given finite time scale.
In the simplest case, one considers a sudden quantum quench, where
$H(t)= H_i$ for $t<0$ and $H(t)=H_f$ for $t>0$.
One then follows the time evolution of a quantum state $|\psi(t)\rangle$,
which is initially in the ground state of $H_i$, and  is then later evolved by $H_f$.

While predicting and classifying the behaviors of $|\psi(t)\rangle$
are daunting tasks in general,
if the dynamics described by the hamiltonian $H_f$ is sufficiently ergodic (chaotic),
it is expected that the late time behaviors of the state $|\psi(t)\rangle$
are well captured by the thermal state --
the state $|\psi(t)\rangle$ {\it thermalizes} at late times
\cite{Page1993, Deutsch1991,Jensen1985,Srednicki1994,Rigol2008,Eisert2015,Kaufman2016}.
For systems which are not chaotic, e.g., integrable systems,
it is expected that the late time behavior of the system after a quantum quench
is characterized by the generalized Gibbs ensemble (GGE)
\cite{Rigol2007}, which has been studied in some solvable models \cite{GGE}.

\paragraph{Quantum quench and thermalization in (1+1)d CFT}
Quantum quenches  and thermalization
in the context of (1+1)-dimensional conformal field theory (CFTs) have been
studied in the literature.
In Ref.\ \cite{Calabrese2006},
thermalization after a global quantum quench
is studied based on the two- and higher point correlation function of local operators.
It is found that when all these local operators fall into the light cone,
the correlation function becomes stationary and equals 
its  value in the thermal ensemble up to
exponentially small corrections.
Later in Ref.\ \cite{Cardy2015},
Cardy calculated the overlap between the reduced density matrix after thermalization (introduced by a global quench) and
that in the  thermal ensemble, and found the overlap is exponentially close to unity.
In the same work, the effect of a  deformation of the initial state and of the CFT hamiltonian
was also studied.
See also Ref.\ \cite{Mandal2015} for a  related discussion.

\paragraph{Setup, purpose and main results of this paper}
In this paper, we will have a further look at thermalization in (1+1)d CFTs,
by focusing on the time evolution of the  entanglement hamiltonian and its spectrum
for a given finite subregion $A$ of the total system.

\begin{figure}[htp]
\begin{center}
\includegraphics[width=5.0in]{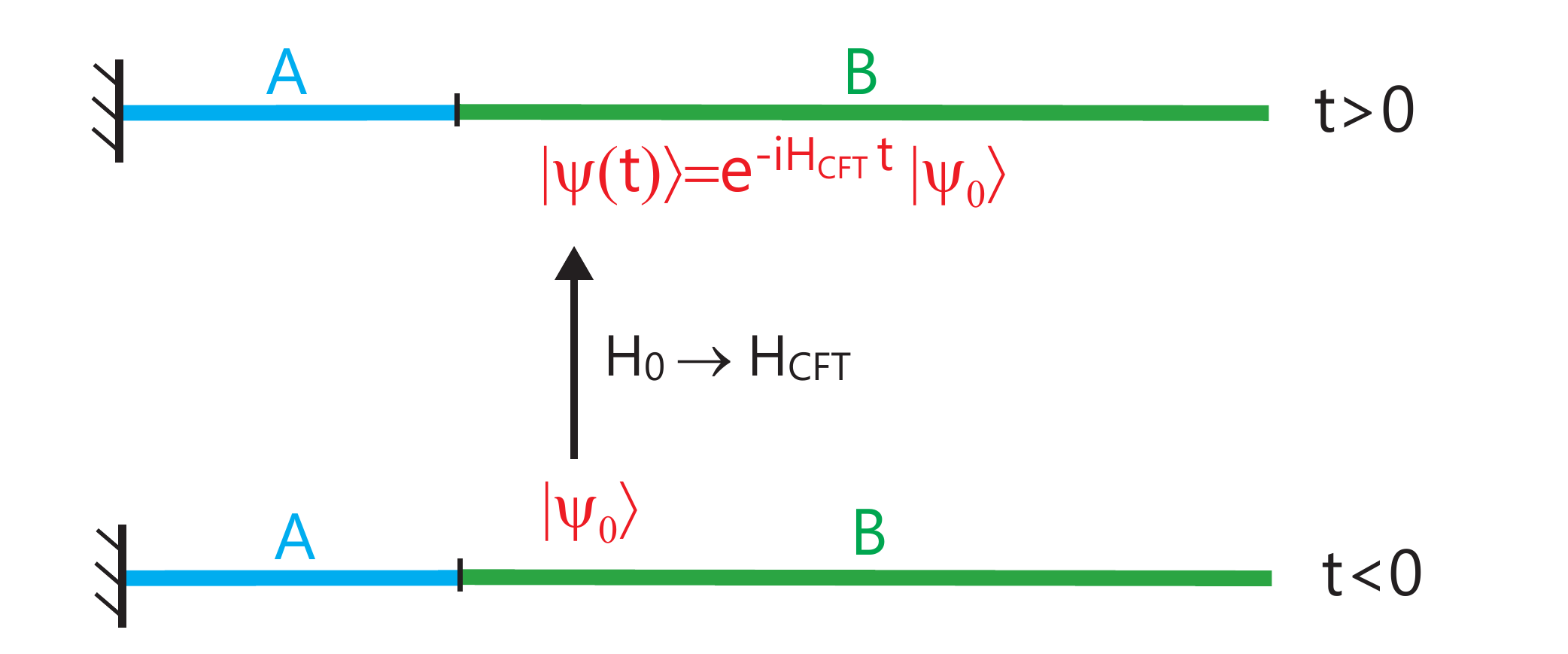}
\caption{Setup of the global quantum quench in this work.
The total system is defined along $[0,\infty)$ and the subsystem $A$ lives on the
finite interval $[0,L]$, at the end of the total semi-infinite system.
One starts from a short-range entangled state
$|\psi_0\rangle$, which may be viewed as the ground state of a gapped hamiltonian $H_0$.
At $t=0$, we switch the hamiltonian to $H_{\text{CFT}}$, and the state $|\psi(t>0)\rangle$ evolves as
$|\psi(t)\rangle=e^{-i H_{\text{CFT}}t}|\psi_0\rangle$.
}\label{Setup}
\end{center}
\end{figure}

To be specific, we will focus on the following setup
for a
 global quantum quench summarized in Fig.\ref{Setup}.
Our system is semi-infinite with a physical boundary at $x=0$.
At time $t=0$, we start from an initial state with
short-range correlations (short-range entangled),
which may be considered
the ground state of a gapped hamiltonian.
We then quench into a CFT;
the state after $t=0$ will then be
 time-evolved with
the hamiltonian $H_{{\rm CFT}}$ which is
the  hamiltonian of a CFT in semi-infinite space
with a boundary at $x=0$.
Of 
 interest
to us is quantum entanglement
between two subregions $A$ and $B$.
We choose subregion $A$ 
to be  a finite interval of length $L$ {\it at the end} of the total system, i.e.
$A=[0, L]$, whereas
the complement
$B=(L, \infty)$
 of $A$
is semi-infinite.
It is expected that the finiteness
 of subsystem $A$ and the semi-infinite size of the complement $B$ (the "bath")
may result in thermalization after 
a quantum quench, as discussed shortly.

In order to diagnose thermalization in this setup,
we will study
(i) the entanglement hamiltonian for the finite subregion $A$,
(ii) its entanglement spectrum,
and, in addition,
(iii) the fictitious {\it  real time}-evolution generated by the entanglement hamiltonian,
which can be visualized by a (Killing) vector field in Minkowski spacetime.
For simplicity, we 
refer to the  latter as the ``modular flow'' (as it is conventionally done).
As we will demonstrate, following the time-dependence of these quantities
provides detailed information
about
the thermalization process.

The main results of this work are summarized as follows:

(i) By choosing the short-range entangled {\it  initial} state as a regularized conformally invariant boundary state,
corresponding to 
a conformal boundary condition that  is the {\it same}  as that characterizing
the {\it physical boundary}  at the  left end of the semi-infinite spatial region at position $x=0$ (see Fig. \ref{Setup}),
we find that in the limit $t\to \infty$ the entanglement hamiltonian and its spectrum for  subsystem $A=[0,L]$
are exactly the same as those in a thermal ensemble  for $H_{CFT}$  with finite temperature  $\beta^{-1}$.
The entanglement spectrum approaches
this   steady  state spectrum  exponentially in time for $t>L$.

(ii) We also study the modular flows generated by the entanglement hamiltonian in Minkowski spacetime.
These flows show different features in the regime where $t<L$ 
as compared to the regime $t>L$.
For $t<L$, these flows evolve in time and tell us how the entanglement propagates; for $t>L$,
these flows become
stationary  and exhibit  the features of the  thermal ensemble, indicating thermalization of
the subsystem.

(iii) We also  study the effect of a generic initial state. When the conformally invariant 
 boundary state describing  the short-range entangled  {\it  initial state}
is different from  that describing  the physical boundary condition at the end  $x=0$ of the semi-infinite space,
 the entanglement spectrum of subsystem  $A=[0,L]$ is not
exactly the same,
in the long time limit $t\to \infty$,
 as that in the  thermal ensemble.
There is an order $\mathcal{O}(1)$
difference in the entanglement entropy $S_A$ as compared to the situation discussion in (i) above.

%

\subsection{Entanglement and entanglement hamiltonian}

For the rest of this section, we will introduce
necessary concepts and terminologies which will be used
throughout the paper.

We first recall the definitions of
the entanglement hamiltonian or modular hamiltonian,
which play
 an important role in understanding 
quantum
entanglement in many-body systems and quantum field theories. For example, the spectrum of the entanglement hamiltonian,
also called entanglement spectrum, is useful for characterizing and classifying gapped quantum many-body states
\cite{RyuHatsugai2006,Haldane2008,Top_1D_0910,Katsura,BauerVidal}. The entanglement hamiltonian is also important for studying the relative entropy
and first law of entanglement \cite{blanco2013}.

Given the  reduced density matrix $\rho_A$ defined for a given subregion $A$,
which encodes all the information 
on the observables localized in the subregion $A$,
the entanglement hamiltonian $K_A$  is defined by
\be
\rho_A=e^{-2\pi K_A}, \quad \text{or} \quad K_A=-\frac{1}{2\pi}\log \rho_A.
\ee
Apparently, the knowledge of  the entanglement hamiltonian $K_A$ is equivalent to that of the  reduced density matrix $\rho_A$.
In particular, the spectrum of $K_A$ determines
all the Renyi entropies and the von-Neumann entropy as follows
\be
S_A^{(n)}=\frac{1}{1-n}\log \text{Tr}_A e^{-2\pi n K_A}, \quad S_A=\lim_{n\to 1}S_A^{(n)}=2\pi \text{Tr}_A\left(K_Ae^{-2\pi K_A}\right).
\ee

Although difficult to obtain in general, there are some specific cases in relativistic quantum field theories where
$K_A$ can be explicitly expressed as an integral of local operators.
One basic example is 
the reduced density matrix for half-space $x_1>0$ of the ground state of a relativistic quantum field theory
in infinite $d$-dimensional (position)  space.
Its  entanglement
hamiltonian can be expressed as
$
K_A=\int_{x_1>0}x_1T_{00}(x)d^{d-1}x
$
\cite{bw,unruh},
where $T_{00}(x)$ denotes the local energy density operator, i.e.  the `time-time' component of the energy momentum tensor.
Other interesting cases include spherical regions in CFTs \cite{chm},
regions in a thermal ensemble in CFTs \cite{Wong2013},
$n$ disjoint intervals for a two dimensional massless Dirac field \cite{Casini0903,Longo0912,Wong1805},
the
Ising chain away from criticality \cite{Peschel1999_2009},
a free-fermion chain with arbitrary filling \cite{Eisler1703},
and a variety of one- and two-dimensional lattice models \cite{ToldinAssaad1804,He1805,EH1807},
\textit{etc}..
To our knowledge, entanglement hamiltonians for time dependent cases were not studied until the most recent work
by Cardy and Tonni \cite{Cardy1608},
where both global and local quantum quenches in
(1+1)d CFTs were considered.
(We will give a brief overview of their approach below.)

\subsection{Quasi particle picture}

A  detailed analysis of the entanglement hamiltonian,
the entanglement spectrum, and the modular flow
in our setup will be presented in the following sections.
Here, we present a simple physical picture
for 
 entanglement propagation
based on the quasi-particle picture (Fig.\ \ref{EPR_Boundary}).
While there is no a priori reason
for the quasi particle picture to hold
for generic interacting CFTs when the central charge is sufficiently large\cite{HartmanJHEP2015},
it allows us to find that most of the results in this paper,
including the time evolution of entanglement hamiltonians and modular flows in Minkowski spacetime,
can be straightforwardly understood in terms of this picture.
In particular, we readily identify
three different time 
regimes  as follows.

At $t=0$, the entanglement between $A$ and $B$
originates mainly from
the region near the entangling point, \textit{i.e.}, from positions $x$ satisfying
 $|x-L|\sim \beta$.
 (Here $\beta$ measures the correlation length of the initial state.).
After $t=0$, quasiparticles are emitted from each point of the system. The entanglement is carried between the
left-moving and right-moving quasiparticles which propagate in opposite directions with the speed of light 
$c=1$.
As shown in Fig.$\,$$\,$\ref{EPR_Boundary},
by focusing on the distribution of quasiparticles in subsystem $A$,
there are mainly three interesting time regimes:

\begin{figure}[htp]
\begin{center}
\includegraphics[width=5.5in]{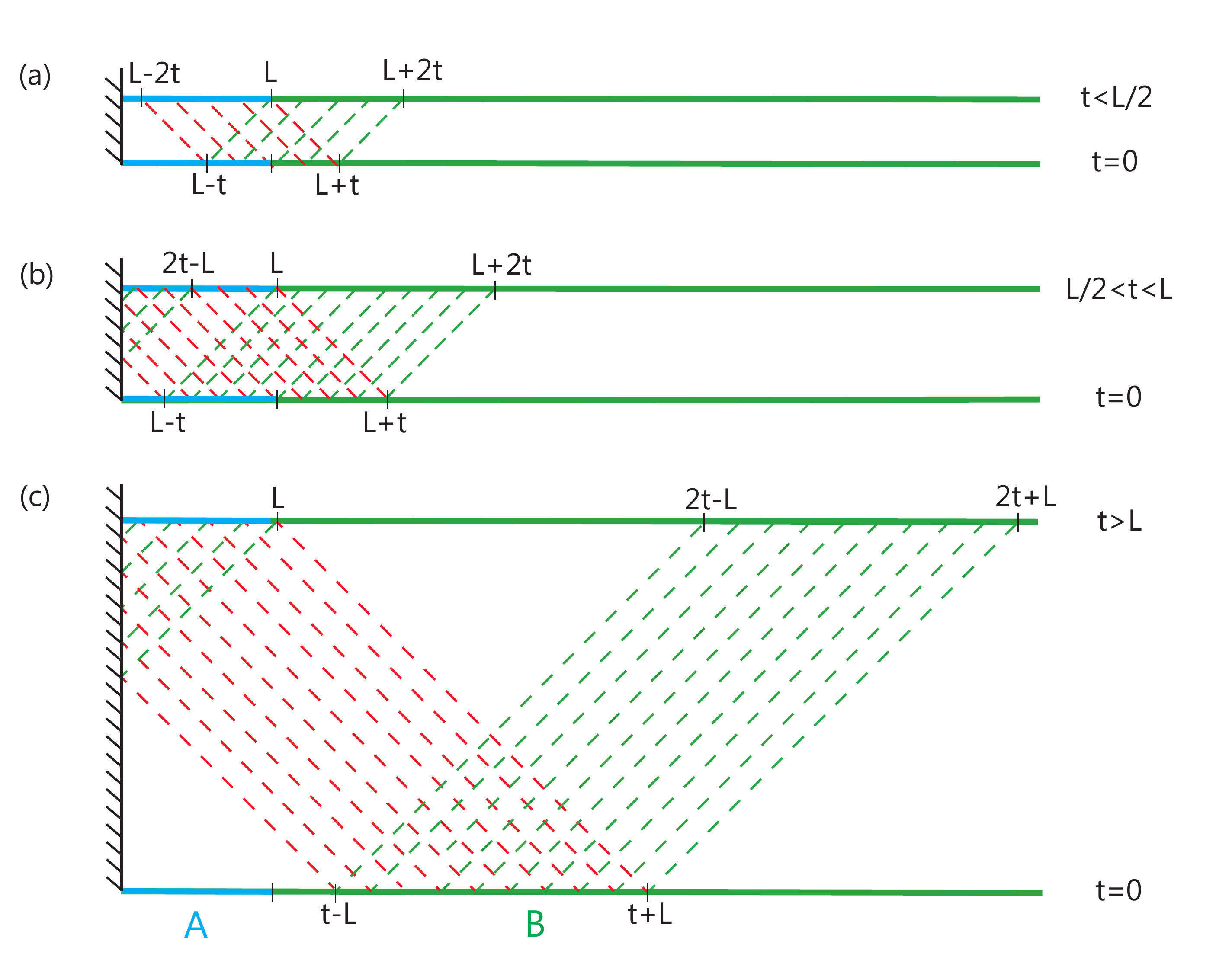}
\caption{Quasiparticle picture describing
the entanglement between $A=[0,L]$ and $B=(L,\infty)$ after a global quantum quench.
Here we focus on the quasiparticles in subsystem $A$.
(a) For $t<L/2$, only the left-moving quasiparticles (red dotted line) in subinterval $(L-2t,L)$ contribute to the
entanglement between $A$ and $B$. These quasiparticles are entangled with the right-moving ones
in subinterval $(L,L+2t)$. (b) For $L/2<t<L$, due
to reflection from the  physical boundary at $x=0$,
both the right-moving quasiparticles (green dotted line) in subinterval $[0,2t-L)$ and the left-moving quasiparticles
in interval $[0,L]$ contribute to the entanglement. They
are entangled with the right-moving quasiparticles in subinterval $(L,L+2t)$. (c) For $t>L$,
the entanglement between $A$ and $B$
is saturated. Both the left-moving and right moving quasiparticles in $[0,L]$ contribute to the entanglement.
These quasiparticles are entangled
with the right-moving ones in subinterval $[2t-L,2t+L]$.
}\label{EPR_Boundary}
\end{center}
\end{figure}

\begin{enumerate}

\item

For $t<L/2$,
only the left-moving quasiparticles distributed in subinterval $[L-2t,L]$ contribute to the entanglement between $A$ and $B$.
At these times $t$, these
quasiparticles
are entangled with those which are right-moving
and located
in subinterval $[L,L+2t]$
of
 subsystem $B$.

\item
 For $L/2<t<L$, both the left-moving quasiparticles in $[0,L]$
and the right-moving quasiparticles in $[0,2t-L]$ contribute to the entanglement \footnote{
The right-moving quasiparticles in subinterval $[0,2t-L]$ come from the left-moving quasiparticles due to  reflection
from the
physical boundary at $x=0$.
}. These quasiparticles are entangled with 
the  right-moving ones in region $[L,L+2t]$
of subsystem $B$.

\item
 For $t>L$, both the left-moving and the right-moving quasiparticles in the whole region of subsystem $A=[0,L]$
contribute to the entanglement, and they are entangled with the right-moving quasiparticles in region $[2t-L,2t+L]$
of subsystem $B$.
In particular, the entanglement entropy
$S_A$  ($S_B$)
 of subsystem $A$ ($B$) is saturated because the number of
quasiparticles that contribute to 
$S_A$  ($S_B$) does no longer
increase.

\end{enumerate}

\subsection{Global quantum quench in (1+1)d CFTs}
\label{Sec: Review Cardy-Tonni}

In CFTs, the entanglement hamiltonian and its  spectrum
can often  be obtained by making use of conformal mapping.
Let us now first  review this approach used  by Cardy and Tonni \cite{Cardy1608},
focusing on global quantum quenches  in an infinite system.

One starts from an initial state $|\psi_0\rangle$ and evolves  it
with a hamiltonian as $e^{-iHt}|\psi_0\rangle$.
Here we choose $H=H_{\text{CFT}}$.
To simplify this problem,
one can choose
an initial state of the form 
 $|\psi_0\rangle=e^{-(\beta/4)H_{\rm{CFT}}}|b\rangle$ where $|b\rangle$
is a conformal boundary state.
The conformal boundary state is a non-normalizable state with
no real-space entanglement \cite{Miyaji}.
Evolving $|b\rangle$
with  a small amount of (imaginary) time $\beta/4$,
introduces  a finite (small) real space entanglement
and the state $e^{-(\beta/4)H_{\rm{CFT}}}|b\rangle$ becomes normalizable \footnote{
The reason we choose $\beta/4$ in the exponential factor is that if we look at  (the expectation value of) the
energy density in this state, it is the same as that in a thermal ensemble at finite temperature $\beta^{-1}$.}.
Physically, the parameter $\beta$ can be interpreted as the correlation length of the initial state. Throughout this work,
we are interested in the limit $L\gg \beta$.
The time dependent density matrix has the form
$
\rho(t)\propto e^{-iHt}e^{-(\beta/4)H_{\rm{CFT}}}|b\rangle\langle b|e^{-(\beta/4)H_{\rm{CFT}}}e^{iHt}.
$
We will work in 
Euclidean spacetime, \textit{i.e.} with
\be\label{rho_global_euclidean}
\rho(\tau)
\propto
e^{-H\tau}e^{-(\beta/4)H_{\rm{CFT}}}|b\rangle\langle b|e^{-(\beta/4)H_{\rm{CFT}}}e^{+H\tau}.
\ee
Quantities such as entanglement entropy, correlation functions of operators and so on can be
evaluated based on $\rho(\tau)$.
To
obtain  the real time evolution,
we simply need to take an analytical continuation $\tau\to it$ in the final
step.

\begin{figure}[htp]
\begin{center}
\includegraphics[width=5.8in]{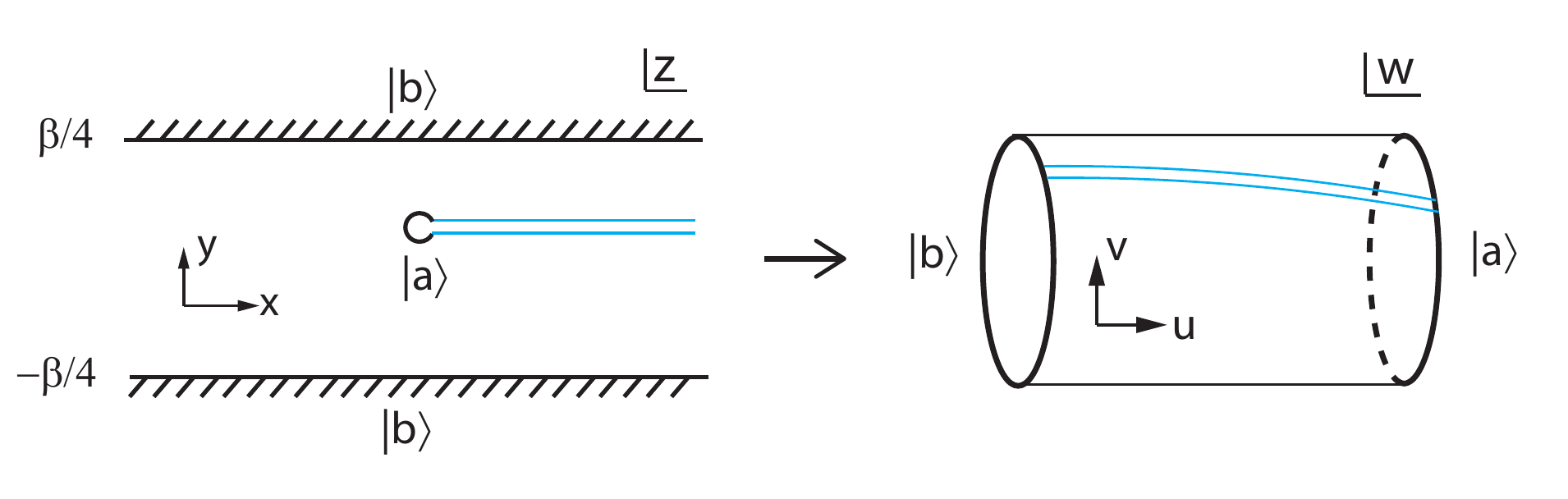}
\end{center}
\caption{
 Euclidean spacetime for $\rho_A(\tau)$ after a global quantum quench, where the semi-infinite subsystem
$A=[0,\infty)$ is part of the infinite real line.
The width of the
strip is $\beta/2$, and
a  branch cut (blue lines) is located along $C=\{i\tau+x,x\ge 0\}$.
The fields living on the upper and lower edges of $C$ correspond to the rows and columns of $\rho_A$.
We remove a small disc at the entangling surface $z_0=i\tau+0$ as regularization, and a conformal boundary
state $|a\rangle$ is imposed along this disc.
With the  conformal mapping $w=f(z)$, the strip is mapped to an annulus, where the time slice $C$ is mapped to $f(C)$
which connects the two edges of the annulus, as shown in the right panel of the Figure. The circumference along
the  $v$-direction is $2\pi$.
}\label{Infinite_Quench}
\end{figure}

In a space-time path integral picture,
$\rho(\tau)$ in Eq.\ (\ref{rho_global_euclidean}) can be represented as a path integral in  a strip of width $\beta/2$,
as shown in
Fig.\ \ref{Infinite_Quench}, where we choose $A=[0,\infty)$ and $B=(-\infty,0)$.
Here, the reduced density matrix 
$\rho_A=\text{Tr}_B\,\rho$  is obtained by sewing together the degrees of freedom in $B$, and then
there is
a branch cut along $C=\{i\tau+x,x\ge 0\}$.
To introduce regularization, we remove a small disc at the entangling point $z_0=i\tau+0$. Then the strip
(with the
small disk  removed)
can be mapped to an annulus in the $w$-plane after a conformal mapping $w=f(z)$. The circumference along
the periodic
$v=\text{Im}\, w$ direction is $2\pi$, and the width of the annulus along the $u=\text{Re}\,w$ direction is denoted by $W$.

Then the entanglement hamiltonian $K_A$,
after the conformal mapping $w=f(z)$,
can be considered as
the generator of translation along the $v$ direction of the annulus.
That is, it can be written in terms of the $vv$ component
of the energy momentum tensor
$T_{\mu\nu}$ as follows
\be
K_A=-\int_{v=\text{const}}T_{vv}du=\int_{f(C)}T(w)dw+\int_{\overline{f(C)}}\overline{T}(\bar{w})d\bar{w},
\ee
where in the second step
we introduce
the holomorphic (antiholomorphic) component of the energy momentum tensor,
$T$ ($\bar{T}$),
and use the fact that
$T_{00}=-T_{vv}=T+\overline{T}$ with the hamiltonian density
$T_{00}$ in Minkowski signature and $T_{vv}$ in Euclidean signature.
Upon mapping back to the $z$-plane, 
$K_A$ becomes
\be\label{KA_half_z}
K_A=\int_{C}\frac{T(z)}{f'(z)}dz+\int_{\overline{C}}\frac{\overline{T}(\bar{z})}{\overline{f'(z)}}d\bar{z},
\ee
where the Schwartzian derivative  term has been ignored since it will be canceled in the calculation of
the  entanglement entropy 
upon 
 introducing
the normalization factor $\text{Tr}\, \rho_A$.
It is noted that the time slices $C$ and $\overline{C}$ in Eq.\ (\ref{KA_half_z})
do not coincide in the quantum quench case \cite{Cardy1608}.

In particular, the entanglement hamiltonian 
for subsystem $A=[0,\infty)$
has
at late times
 the approximate form
\be\label{KA_Cardy}
K_A(t)\simeq \frac{\beta}{2\pi}\int_0^{2t}T(x,t)dx,
\ee
where $T=(T_{00}+T_{10})/2$ is the
energy-momentum  tensor for only the right-movers,
and involves both the hamiltonian density $T_{00}$ and the momentum density $T_{10}$.
This result may be understood based on the quasi-particle picture in that only the right-moving quasiparticles
in subsystem $A$ contribute to the entanglement (they are entangled with the left-moving ones in $B$).

The spectrum of $K_A(t)$ can also be obtained based on the knowledge of boundary CFT.
The eigenvalues of $K_A$ are, up to a global shift,
given by $\pi(\Delta_j-c/24)/W$ with degeneracies $d_j$ and central charge $c$.
Here $\Delta_j$ are scaling dimensions of 
 boundary operators.
Then the spacing between levels  of the  entanglement spectrum may be expressed as
\be
E_i-E_j=\frac{\pi (\Delta_i-\Delta_j)}{W}.
\ee
It can be
seen
that the lowest eigenvalue $E_0$ has the form \cite{Cardy1608}
\be
E_0\simeq
\frac{c}{12}W+\log \langle a|0\rangle+\log \langle b|0\rangle.
\ee

To obtain the Renyi or von Neumann entropy, we need to evaluate the partition function $Z_1$ ($Z_n$)
on the annulus in Fig.$\,$$\,$\ref{Infinite_Quench} below,
with circumference $2\pi$  ($2n\pi$). It can be shown that, in the limit $W\gg 1$, one has
\be\label{Tr_rhoA_N}
\frac{\text{Tr}(\rho_A^n)}
{(\text{Tr}\rho_A)^n}=\frac{Z_n}{(Z_1)^n}\simeq \frac{\langle a|0\rangle\langle 0|b\rangle \tilde{q}^{-c/24n}}
{(\langle a|0\rangle\langle 0|b\rangle)^n \tilde{q}^{-cn/24}}, \quad \text{with}\quad \tilde{q}=e^{-2W},
\ee
where $|0\rangle$ denotes the ground state of the CFT.
Then one can obtain the Renyi entropy 
($S^{(n)}_A$)
and von Neumann entropy 
($S_A$)
as follows
\be\label{RvN_Entropy}
S^{(n)}_A=\frac{1}{1-n}\frac{\text{Tr}(\rho_A^n)}
{(\text{Tr}\rho_A)^n}
\simeq \frac{c}{12}\left(1+\frac{1}{n}\right)W-g_a-g_b, \quad S_A\simeq \frac{c}{6}W-g_a-g_b,
\ee
where $g_{a,b}=-\log\langle a,b|0\rangle$ are the Affleck-Ludwig boundary entropies \cite{Affleck-Ludwig}.

Observe the lack of thermalization in
the entanglement hamiltonian \eqref{KA_Cardy},
as it depends both on the hamiltonian and the momentum,
which is due to the fact that $A$ is semi-infinite.
Therefore, it is natural to ask the following questions:
Can we study the time evolution of the entanglement hamiltonian for a finite subsystem during thermalization?
If so, how does the entanglement spectrum evolve in time during
this process? In particular,
how does the entanglement spectrum converge to the saturated
spectrum 
of  a thermal ensemble in the long-time limit?
Are there any quantities visualizing how the entanglement propagates, and how the subsystem is thermalized?
To answer these questions, in this paper we are interested in the time evolution of  the entanglement hamiltonian
and related quantities for a 
{\it finite interval, located at the end of a semi-infinite system}, after a global quantum quench
(see Fig.$\,$$\,$\ref{Setup}).
The reason we choose this configuration is because in this case
the corresponding path integral representation
of the reduced density matrix can be mapped to an annulus and can  then be treated analytically:  Then one can use Cardy-Tonni's approach to
study analytically  the behavior of entanglement hamiltonian/spectrum in this case.

\section{Time evolution of entanglement hamiltonian}
\label{Sec: Semi-infinite}

\subsection{Conformal mapping}

\begin{figure}[htp]
\begin{center}
\includegraphics[width=4.8in]{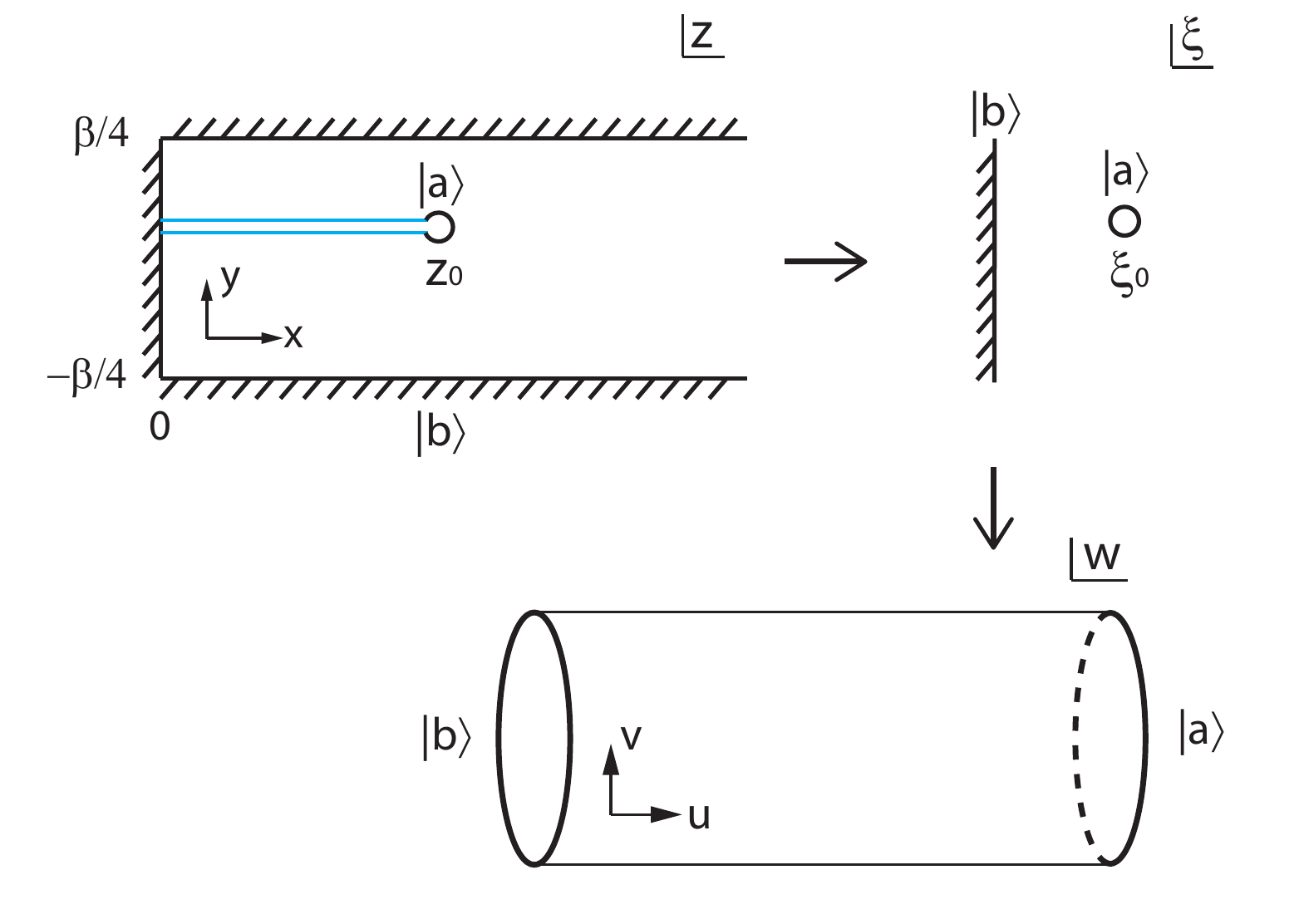}
\end{center}
\caption{Euclidean spacetime for $\rho_A$
after a global quantum quench,  where the finite interval $A=[0,L]$ is at the end of
a semi-infinite system $A\cup B=[0,\infty)$.
The width of the  half-strip is $\beta/2$, and the branch cut (blue lines) is along $C=\{i\tau+x,0\le x\le L\}$.
We remove a small disc at the entangling surface $z_0=i\tau+L$ as regularization.
Conformal boundary conditions $|a\rangle$ and $|b\rangle$ are imposed, respectively,
 at the small circle at $z_0=i\tau+L$,
and at  the vertical and horizontal boundaries along $x=0$ and $y=\pm\frac{\beta}{4}$.
The half strip is mapped to
the right half plane (RHP) after the first conformal mapping $\xi=\xi(z)$, and further mapped to an annulus
after the second conformal mapping $w=w(\xi)$.
Here, we do not show explicitly the  mapping of $C$ between  the $\xi$- and the $z$-plane.
}\label{ConformalMap_Half}
\end{figure}

Shown in Fig.$\,$$\,$\ref{ConformalMap_Half} is our setup for a global quantum quench in a semi-infinite system.
As compared to the case in Fig.$\,$$\,$\ref{Infinite_Quench}, 
we now have a {\it  physical boundary}  at position $x=0$, where
we will impose a conformal boundary condition. In  the space-(Euclidean)time picture of
Fig.$\,$$\,$\ref{ConformalMap_Half}, this (spatial)  boundary condition
appears at the vertical boundary $0+i y$ ($-\beta/4 \leq y \leq +\beta/4$).
In general, this boundary condition can be different from the {\it  initial}  condition which is imposed in
the space-(Euclidean)time picture of
Fig.$\,$$\,$\ref{ConformalMap_Half} at 
the horizontal boundary
$x\pm i\beta/4$ ($0\leq x <\infty$), which is also represented
(as mentioned) by a conformal boundary condition.
When these two boundary conditions are different,  one needs to consider boundary condition changing
operators located at points where two different boundary conditions meet.
For simplicity, we first assume
that these two boundary conditions are the  same,
 and we denote
the corresponding conformal boundary state
by $|b\rangle$.
In addition, in order  to take into account 
regularization, one removes a small disc at the entangling
point located at $i\tau+L$, and imposes a conformal boundary condition $|a\rangle$ at the boundary of this (removed) disk
(see Fig.$\,$$\,$\ref{ConformalMap_Half}).

\begin{figure}[htp]
\begin{center}
\includegraphics[width=6.0in]{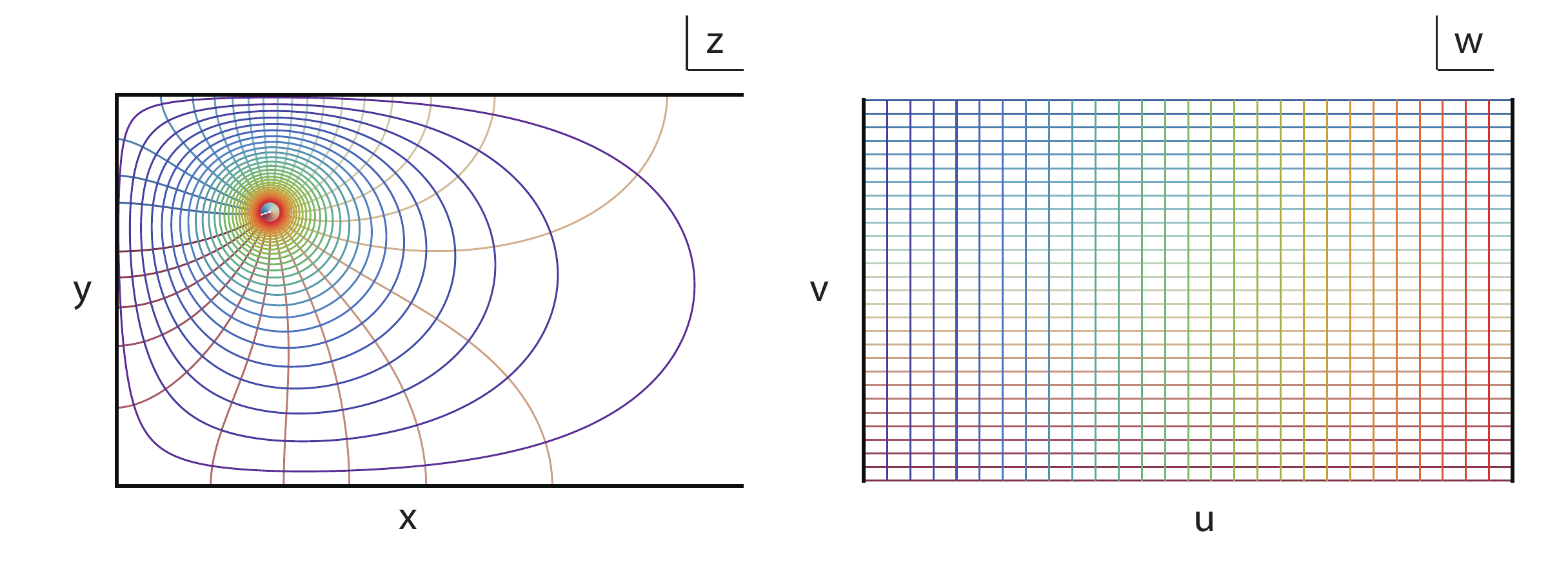}
\end{center}
\caption{Euclidean spacetime for a global quench of a semi-infinite system, where
$A=[0,L]$ and $A\cup B=[0,\infty)$. The vertical width of the half strip is $\beta/2$.
The curves in the $z$-plane in the left panel correspond to constant-$u$ and
constant-$v$ curves in the  $w$-plane in the right panel,
with $w=u+iv$, and $v\in[-\pi,\pi)$.
}\label{halfline}
\end{figure}

It is noted that the Euclidean spacetime for this case is conformally equivalent to an annulus, and therefore one
can use the method in Ref.\ \cite{Cardy1608}.
As shown in Fig.$\,$$\,$\ref{ConformalMap_Half}, one can map\footnote{See Appendix A in
Ref.$\,$\cite{Cardy1608} for more details on the cases with an external boundary.}
 the half strip in the $z$-plane (with the small
disc at the entangling point $i\tau+L$ removed)  to an annulus in the  $w$-plane,
by considering the following two-step conformal mapping $w=f(z)$,
\be\label{two-step-half-line}
\left\{
\begin{split}
\xi(z)=&\sinh \left(\frac{2\pi z}{\beta}\right), \\
w(\xi)=&-\log \left[\left(\frac{1+\overline{\xi}_0}{1+\xi_0}\right)\cdot \frac{\xi-\xi_0}{\xi+\overline{\xi}_0}\right],
\end{split}
\right.
\ee
where
\be
\xi=\xi(z) \quad\text{and}\quad
\xi_0=\xi(z_0),\quad \text{with}\quad z_0=i\tau+L.
\ee
The conformal mapping $\xi(z)$ in the first step maps the semi-infinite strip in the $z$-plane
to the right half plane (RHP), namely
$\rm{Re}(\xi)\ge0$, in the $\xi$-plane  (see Fig.$\,$$\,$\ref{ConformalMap_Half}).
The small disc around the entangling surface $z_0=i\tau+L$ in the $z$-plane is mapped to a small disc around
$\xi_0=\xi(z_0)$ in the RHP.
Then the second conformal mapping $w(\xi)$ sends the RHP with a small disc
at $\xi_0$ removed to an annulus in the  $w$-plane,
where we write $w=u+iv$ with $u$ and $v$ real. After this two-step conformal mapping, the two boundaries labeled by
$|a\rangle$ and $|b\rangle$  in the $z$-plane are mapped to the two boundaries
(edges) 
of  the annulus in the  $w$-plane,
described by
 $\{u=f(i\tau+\epsilon),0\le v<2\pi\}$ and
$\{u=f(i\tau+L-\epsilon), 0\le v<2\pi\}$, respectively.
In
 Fig.$\,$$\,$$\,$\ref{halfline} we
show
 the constant-$u$ and constant-$v$ flows in the $z$-plane
and the  $w$-plane, respectively.
In particular, in the  $w$-plane, these are straight lines, where 
 $u$ runs from $f(\epsilon+i\tau)$ to $f(L-\epsilon+i\tau)$, and $v$ runs from $-\pi$ to $\pi$. 
In the path integral description of $\text{Tr}\,\rho_A$,
 the two segments along $v=-\pi$ and $v=\pi$ are identified.

\subsection{Time evolution of entanglement hamiltonian }

\subsubsection{Entanglement hamiltonian for subsystem $[0,L]$}$\,$

Based on Eq.\ (\ref{KA_half_z}) and  the conformal mapping in Eq.\ (\ref{two-step-half-line}),
it is straightforward to obtain the entanglement hamiltonian $K_A(t)$ for subsystem $A$ as
follows (after analytical continuation $\tau\to it$):
\begin{equation}\label{KAtotal_half}
\begin{split}
K_A(t)
=&\frac{\beta}{\pi}\int_{L}^{0}
\frac
{
\sinh\left[\frac{\pi (x-L)}{\beta}\right]
\cosh\left[\frac{\pi\left(x-2t+L\right)}{\beta}\right]
\sinh\left[\frac{\pi(x+L)}{\beta}\right]
\cosh\left[\frac{\pi(x-2t-L)}{\beta}\right]
}
{\cosh\left(\frac{2\pi}{\beta}t\right)\sinh\left(\frac{2\pi}{\beta}L\right)
\cosh\left[
\frac{2\pi}{\beta}\left(x-t\right)
\right]}
T(x,t)dx\\
\rule{0pt}{.8cm}
&+
\frac{\beta}{\pi}\int_{L}^{0}
\frac
{
\sinh\left[\frac{\pi (x-L) }{\beta}\right]
\cosh\left[\frac{\pi \left(x+2t+L\right)}{\beta}\right]
\sinh\left[\frac{\pi (x+L)}{\beta}\right]
\cosh\left[\frac{\pi (x+2t-L)}{\beta}\right]
}
{\cosh\left(\frac{2\pi t}{\beta}\right)\sinh\left(\frac{2\pi L}{\beta}\right)
\cosh\left[
\frac{2\pi \left(x+t\right)}{\beta}
\right]}
\overline{T}(x,t)dx.
\end{split}
\end{equation}
Let us check the behavior of $K_A(t)$ at $t=0$ first. There are
the  following two interesting cases:

(i) $t=0, \beta\to \infty$.
In this case, the width of the strip in Fig.$\,$$\,$\ref{ConformalMap_Half} goes to infinity, and the initial state at $t=0$
is no longer a short-range entangled state.
 It corresponds to the ground state of a CFT with
a physical boundary at $x=0$. After some simple algebra,  Eq.\ (\ref{KAtotal_half}) can be simplified
to
\begin{equation}\label{KA_infinite_beta_I}
K_A(t=0,\beta\to \infty)\simeq \int_{0}^L
\frac{L^2-x^2}{2L}T_{00}(x)dx,
\end{equation}
which agrees with the (known)  result for the entanglement hamiltonian of
 a finite interval at the end of a semi-infinite CFT.

(ii) $t=0, L\gg \beta$.
In this case, $L$ is much larger than the correlation length of the initial state.
It is straightforward to find that
\begin{eqnarray}
K_A(t=0,L\gg \beta)
=&
\frac{\beta}{2\pi}\int_{0}^L
\sinh\left[\frac{2\pi}{\beta}(L-x)\right]T_{00}(x)dx.
\end{eqnarray}
Here, the
 contribution to the entanglement between $A$ and $B$ mainly comes from the
region near the entangling point, \textit{i.e.},
$(L-x)\sim \mathcal{O}(\beta)$,  as appropriate
for  a short-range entangled state.
For $(L-x)\gg \beta$, the entanglement hamiltonian becomes exponentially large, and its contribution
to the entanglement becomes exponentially suppressed, as expected.

Now let us focus on the time evolution of $K_A(t)$ for $t>0$. $K_A(t)$
shows different behaviors in different 
time-regimes:  (See Appendix \ref{Subsection: EH} for details.)

\begin{equation}\label{KA_3_region}
K_A(t)\simeq\left\{
\begin{split}
&\frac{\beta}{2\pi}\int_{L-2t}^L\overline{T}(x,t)dx, \quad &t<L/2,\\
&\frac{\beta}{2\pi}\int_0^{2t-L}T_{00}(x,t)dx+\frac{\beta}{2\pi}\int_{2t-L}^L\overline{T}(x,t)dx,\quad  &L/2<t<L,\\
&\frac{\beta}{2\pi}\int_0^L  T_{00}(x,t)dx, \quad & t>L.
\end{split}
\right.
\end{equation}
Here we have ignored the interesting contributions close to the entangling point,
\textit{i.e.}, contributions coming from regions $L-x\sim O(\beta)$ [see Eqs.\ (\ref{KAI_i})-(\ref{KAI_ii}) in Appendix].
$T$, $\overline{T}$ and $T_{00}$ are related
via
 $T=(T_{00}+T_{10})/2$ and $\overline{T}=(T_{00}-T_{10})/2$,
where $T$ ($\overline{T}$) is the energy-momentum tensor for the right (left) movers.
The behavior of $K_A(t)$ in Eq.\ (\ref{KA_3_region}) may be understood as follows:

(i) $t<L/2$.
Only the energy momentum tensor for the \textit{left} movers, namely $\overline{T}(x,t)$,  appears in $K_A(t)$.
This can be easily understood based on Fig.$\,$$\,$\ref{EPR_Boundary}(a), where only the left-moving quasiparticles
in interval $A$ contribute to the entanglement between $A$ and $B$.
In particular, these quasiparticles are distributed in subinterval $[L-2t, L]$, corresponding to
the integral
 $\int_{L-2t}^L\cdots dx$ in the expression for  $K_A(t<L/2)$ in Eq. (\ref{KA_3_region}).

(ii) $L/2<t<L$. Here, 
$K_A(t)$  can be rewritten as
\begin{equation}
\begin{split}
K_A\left(L/2<t<L\right)
\simeq \frac{\beta}{2\pi}\int_0^{2t-L}T(x,t)dx+\frac{\beta}{2\pi}\int_{0}^L\overline{T}(x,t)dx.
\end{split}
\end{equation}
That is, the right movers $T(x,t)$ start to contribute to $K_A(t)$.
The
intervals over which the corresponding integrals extend,
i.e. $\int_0^{2t-L}T(x,t)dx$ and $\int_{0}^L\overline{T}(x,t)dx$ respectively,
also agree with the physical picture in Fig.$\,$$\,$\ref{EPR_Boundary}$\,$(b) which says  that the right-moving quasiparticles in
$[0,2t-L]$ and the  left-moving quasiparticles in $[0,L]$ contribute to the entanglement between $A$ and $B$.

(iii) $t>L$.
Only the hamiltonian density $T_{00}(x,t)$ appears in $K_A(t)$.
Considering that $T_{00}(x,t)=T(x,t)+\overline{T}(x,t)$, the term proportional to $\int_0^L  T_{00}(x,t)dx$
agrees with the physical picture in Fig.$\,$$\,$\ref{EPR_Boundary}$\,$(c) which says that both the left-moving and right-moving
quasiparticles distributed in $[0,L]$ contribute to the entanglement.

Note that we have ignored the contributions near the entangling point when evaluating $K_A(t)$ in Eq.\ (\ref{KA_3_region}).
In particular,
in the long time  limit $t\to \infty$, one obtains  the following expression when keeping all contributions
\be
K_A(t\to\infty)=\frac{\beta}{\pi}\int_{0}^L \frac{
\sinh[\pi(L-x)/\beta]\sinh[\pi(L+x)/\beta]
}{
\sinh(2\pi L/\beta)
}T_{00}(x,t)dx,
\ee
which has the same form as  that 
of
 the entanglement hamiltonian
 in 
the  thermal ensemble [see Eq.\ (\ref{FiniteT_KA_semi_infinite})].
In fact, as shown
below,
 for $t\to\infty$  the spectrum of $K_A(t)$ is exactly the same as that in the  thermal ensemble. This indicates that
 in the long time limit $t\to\infty$
 the reduced density matrix $\rho_A(t)$ 
is exactly the same as the reduced density matrix $\rho_A(\beta)$  
in the  thermal ensemble at finite temperature $\beta^{-1}$.

\subsubsection{Entanglement hamiltonian for subsystem $(L,\infty)$}$\,$

To obtain the entanglement hamiltonian $K_B(t)$ for subsystem $B = (L, \infty)$, we simply need to replace the
path $C=\{i\tau+x, 0\le x\le L\}$ with
the path  $C=\{i\tau+x, x>L\}$ in Eq.\ (\ref{KA_half_z})
[see also Fig.$\,$$\,$\ref{ConformalMap_Half}],
and therefore change the
interval over which  the integral 
 in Eq.\ (\ref{KAtotal_half})
is taken, from
$\int_L^0 \cdots dx \to \int_{L}^{\infty}\cdots dx$.
After some simple algebra, one
finds
\begin{equation}\label{KB_2_region}
K_B(t)\simeq
\left\{
\begin{split}
&\frac{\beta}{2\pi}\int_L^{2t+L}T(x,t)dx,\quad t<L,\\
&\frac{\beta}{2\pi}\int_{2t-L}^{2t+L}T(x,t)dx, \quad t>L,
\end{split}
\right.
\end{equation}
where, again, we have ignored the contributions close to the entangling point,
\textit{i.e.} from the region 
$x-L\sim O(\beta)$.
One interesting feature
in the expressions for
 $K_B(t)$
above  is that only the energy momentum tensor for the
right movers, namely $T(x,t)$, appears in $K_B(t)$.
This can be easily understood based on the quasi-particle picture in Fig.\
\ref{EPR_Boundary},
where only the right movers distributed in $[L,2t+L]$ for $t<L$
 (and distributed in $[2t-L,2t+L]$ for $t>L$) in subsystem $B$ contribute to the
entanglement between $B$ and $A$.
We emphasize the difference between Eq.\ (\ref{KB_2_region}), and Eq.\
(\ref{KA_Cardy})
valid
for subsystem $[0,\infty)$ in the infinite system $(-\infty, \infty)$ \cite{Cardy1608}.
For $t<L$, the result in Eq.\ (\ref{KB_2_region}) agrees with Eq.\ (\ref{KA_Cardy}) by setting $L=0$.
For $t>L$, however, the
interval over which the integral 
 in Eq.\ (\ref{KB_2_region}) is taken  is $[2t-L,2t+L]$ with a constant width $2L$,
which is different from the
 interval $[0,2t]$ appearing in Eq.\ (\ref{KA_Cardy}) which grows linearly in $t$.
This is because the reservoir for the entanglement Hamiltonian  in Eq.\ (\ref{KB_2_region}) for region
 $B=[L,\infty)$ 
is region $A=[0,L]$ which is finite, 
while
the reservoir
of the entanglement hamiltonian in Eq.\ (\ref{KA_Cardy})  for  region
$A=[0,\infty)$ is region  $B=(-\infty,0)$, which is infinite.

In addition, note that in the long time limit $t\to\infty$, the entanglement hamiltonian  $K_B(t)$ in Eq.\ (\ref{KB_2_region}) can never
approach that in a thermal ensemble. In other words,
 subsystem $B$ can not thermalize,
because it is of infinite spatial extent. 

\section{Time evolution of entanglement spectrum and entanglement entropy}
\label{Sec: Semi-infinite Sepctrum}

Here we use the method briefly reviewed in Sec.\ \ref{Sec: Review Cardy-Tonni} to
study the time evolution of the  entanglement spectrum and of   the entanglement entropy.
By defining
\be
\mathcal{W}=f(i\tau+L-\epsilon)-f(i\tau),
\ee
where $f(z)$ is the conformal mapping in Eq.\ (\ref{two-step-half-line}),
the width $W$ of the annulus (see Fig.$\,$\ref{ConformalMap_Half}) can be expressed as
\begin{equation}
W=\text{Re}(\mathcal{W})=\frac{1}{2}\left(\mathcal{W}+\overline{\mathcal{W}}\right).
\end{equation}
After some straightforward algebra, one 
finds that  $\mathcal{W}$ 
  has the explicit form
\be\label{W_Euclidean_caseI}
\mathcal{W}=
\log
\left\{
\frac
{\cosh\left[\frac{2\pi}{\beta}\cdot\frac{i2\tau+L}{2}\right]}
{ \cosh\left[\frac{2\pi}{\beta}\cdot\frac{i2\tau-L}{2}\right]}
\cdot
\frac
{\sinh\left(\frac{2\pi}{\beta}\cdot\frac{2L-\epsilon}{2}\right)\cdot \cosh\left(\frac{2\pi}{\beta}\cdot\frac{i2\tau-\epsilon}{2}\right)}
{\sinh\left(\frac{2\pi}{\beta}\cdot\frac{\epsilon}{2}\right)\cdot \cosh\left(\frac{2\pi}{\beta}\cdot\frac{i2\tau+2L-\epsilon}{2}\right)}
\right\}.
\ee
Upon analytical continuation
to real time, $\tau\to it$, one obtains
\begin{equation}\label{WA_I}
W=
\frac{1}{2}(\mathcal{W}+\overline{\mathcal{W}})
=
\log
\left\{
\frac
{2 \sinh\left[\frac{2\pi }{\beta}(L-\frac{\epsilon}{2})\right]\cdot \cosh\left(\frac{2\pi t}{\beta} \right)}
{\sinh\left(\frac{2\pi}{\beta}\cdot \frac{\epsilon}{2}\right)\cdot \sqrt{2\cosh\left(\frac{2\pi}{\beta}\cdot 2L\right)+2\cosh\left(\frac{2\pi}{\beta}\cdot 2t\right)}}
\right\}.
\end{equation}
By further considering the limit $L, t\gg \beta\gg \epsilon$,  this expression for $W$ can be simplified as
\begin{equation}\label{SA(t)_I}
W\simeq
\left\{
\begin{split}
&\log\left(\frac{\beta}{2\pi\epsilon}\right)+\frac{2\pi}{\beta}t, \quad & t<L,\\
&\log\left(\frac{\beta}{2\pi\epsilon}\right)+\frac{2\pi}{\beta}L, \quad & t>L,
\end{split}
\right.
\quad
\Rightarrow\quad
S_A(t)\simeq
\left\{
\begin{split}
&\frac{\pi c}{3\beta} t,\quad &t<L,\\
&\frac{\pi c}{3\beta} L,\quad &t>L,
\end{split}
\right.
\end{equation}
where in the second step we have used Eq.\ (\ref{RvN_Entropy}), and 
have only kept  the leading term in $t$ or $L$.
For $t<L$, $S_A\simeq (\pi c/3\beta)t$, \textit{i.e.}, the entanglement entropy grows linearly in time.
For $t>L$, both $W$ and  the entanglement entropy $S_A(t)$ are saturated. They are the same as those
displayed in 
Eqs.\ (\ref{Thermal_W}) and (\ref{Thermal_vN_Entropy}) for the  thermal ensemble, to 
 leading order in large quantities.

As just mentioned,
the results in Eq.\ (\ref{SA(t)_I}) are approximated  in that only the leading order in $t$ or $L$ has been kept.
In fact, in the limit $t\to \infty$,
one can find the \textit{exact} expression of $W$ based on Eq.\ (\ref{WA_I}),
which reads as follows
\be
W(t\to \infty)=\log\frac{
\sinh[\pi(2L-\epsilon)/\beta]
}
{\sinh(\pi \epsilon/\beta)}=:W_{\text{thermal}}.
\ee
This expression
 is exactly the same as that for the  thermal ensemble, displayed in Eq.\ (\ref{Thermal_half_W}), indicating
that the long time limit of the  reduced density matrix $\rho_A(t\to \infty)$ is \textit{indistinguishable}
from the reduced density matrix in the  thermal ensemble
at finite
temperature $\beta^{-1}$.
Then the level spacing of the  entanglement spectrum has the following form
\be
E_i-E_j=\frac{\pi (\Delta_i-\Delta_j)}{W_\text{thermal}}, \quad\text{with } t\to \infty.
\ee

It is interesting to check how the width $W(t)$ in Eq.\ (\ref{WA_I})
approaches the saturated long time value $W_{\text{thermal}}$ as a function of time.
For $t-L\gg \beta$, by expanding
$W$ to the term in $t$, it is straightforward to obtain
\be
W(t>L)\simeq W_{\text{thermal}}-\frac{1}{2}e^{-\frac{4\pi}{\beta}(t-L)}.
\ee
Therefore, for $t-L\gg\beta$, one
obtains the following behavior of
the spacing of the entanglement spectrum 
\be
E_i(t)-E_j(t)=
\frac{\pi (\Delta_i-\Delta_j)}{W}\simeq
\frac{\pi(\Delta_i-\Delta_j)}{W_{\text{thermal}}}
\left[
1+\frac{1}{2W_{\text{thermal}}}
e^{-\frac{4\pi}{\beta}(t-L)}
\right],\quad \text{with }t-L\gg\beta.
\ee
That is, the spacing of  the entanglement spectrum converges
 exponentially in time
 to
its  saturated  long time value 
$\pi (\Delta_i-\Delta_j)/W_{\text{thermal}}$.

It is also worth checking the behavior of $\mathcal{W}$ and $\overline{\mathcal{W}}$ respectively
after analytical continuation $\tau\to it$.
In the region $t<L/2$, based on Eq.\ (\ref{W_Euclidean_caseI}) and its complex conjugate $\overline{\mathcal{W}}$, one has
\be\label{W_01}
\left\{
\begin{split}
\mathcal{W}|_{\tau=it}&\simeq  \log\left(\frac{\beta}{2\pi \epsilon}\right), \\
\overline{\mathcal{W}}|_{\tau=it}&\simeq\log\left(
\frac{\beta}{2\pi\epsilon}
\right)+
\frac{2\pi}{\beta}\cdot 2t. \\
\end{split}
\right.
\ee
The entanglement entropy mainly comes from $\overline{\mathcal{W}}|_{\tau=it}$, \textit{i.e.}, from  the left movers.
This agrees with Eq.\ (\ref{KA_3_region}) which says that only the left movers $\overline{T}(x,t)$ appear in the entanglement
hamiltonian $K_A(t<L/2)$.
It is
 remarkable that the factor $2t$ in the expression for  $\overline{W}|_{\tau\to it}$ corresponds to the length
of the interval  $[L-2t, L]$ for the left-moving quasiparticles in Fig.$\,$$\,$\ref{EPR_Boundary}(a).

In the region $L/2<t<L$, one
obtains
\be\label{W_02}
\left\{
\begin{split}
\mathcal{W}|_{\tau=it}&\simeq  \log\left(\frac{\beta}{2\pi \epsilon}\right)+\frac{2\pi}{\beta}(2t-L), \\
\overline{\mathcal{W}}|_{\tau=it}&\simeq\log\left(
\frac{\beta}{2\pi\epsilon}\right)+\frac{2\pi}{\beta}L. \\
\end{split}
\right.
\ee
Now, both $\mathcal{W}|_{\tau\to it}$ and $\overline{\mathcal{W}}|_{\tau\to it}$ contribute to the entanglement entropy $S_A(t)$.
In particular, the factor $(2t-L)$ in the expression for $\mathcal{W}|_{\tau\to it}$
corresponds to the length of the interval  $[0,2t-L]$ occupied by the right-moving quasiparticles,
and the factor $L$ in the expression for $\overline{\mathcal{W}}|_{\tau\to it}$ corresponds to the length of the interval $[0,L]$ for
the left-moving quasiparticles in Fig.$\,$$\,$\ref{EPR_Boundary}(b).

For $t>L$, one
obtains
\be
\left\{
\begin{split}
\mathcal{W}|_{\tau=it}&\simeq  \log\left(\frac{\beta}{2\pi \epsilon}\right)  +\frac{2\pi}{\beta}L, \\
\overline{\mathcal{W}}|_{\tau=it}&\simeq\log\left(
\frac{\beta}{2\pi\epsilon}
\right)+\frac{2\pi}{\beta}L. \\
\end{split}
\right.
\ee
The contributions  of $\mathcal{W}|_{\tau\to it}$ and $\overline{\mathcal{W}}|_{\tau\to it}$ to
the entanglement entropy $S_A(t)$ are the same. The factor $L$ in $\mathcal{W}|_{\tau\to it}$
and $\overline{\mathcal{W}}|_{\tau\to it}$ agrees with
the length of $[0,L]$ which is occupied by both left-moving and right-moving quasiparticles in Fig.$\,$$\,$\ref{EPR_Boundary}(c).

\section{Modular flows in Minkowski spacetime}
\label{Sec: constant u flows}
In this section, we study the modular flow,
the fictitious  real-time-evolution generated by the entanglement hamiltonian; it
is represented by a (Killing) vector field in Minkowski spacetime.
In terms of the $w$-coordinate, this is the flow generated by keeping $u$ constant
and varying $v$.
Our motivation to study the constant-$u$ flows is very straightforward:
In the previous part, we have seen that the entanglement entropy $S_A$ is proportional to
the width $W$ of annulus in  the $w$-plane [see Eq.~(\ref{RvN_Entropy})]. Note that $W$ measures the range of
the variable  $u=\text{Re}\,w$
in the  $w$-plane. Therefore, the constant-$u$ flows with $u_{\text{max}}-u_{\text{min}}=W$
should provide us information 
about the entanglement between $A$ and $B$.
Moreover, in Sec.\ref{Sec: Semi-infinite}, 
we have seen that the time evolution of the entanglement entropy and entanglement hamiltonian can be well understood
in terms of
the quasi-particle picture. Based on the above observations, we expect there should be a correspondence
between the patterns of modular flows and the quasi-particle picture, as studied in detail in the following.

\begin{figure}[htp]
\begin{center}
\includegraphics[width=4.2in]{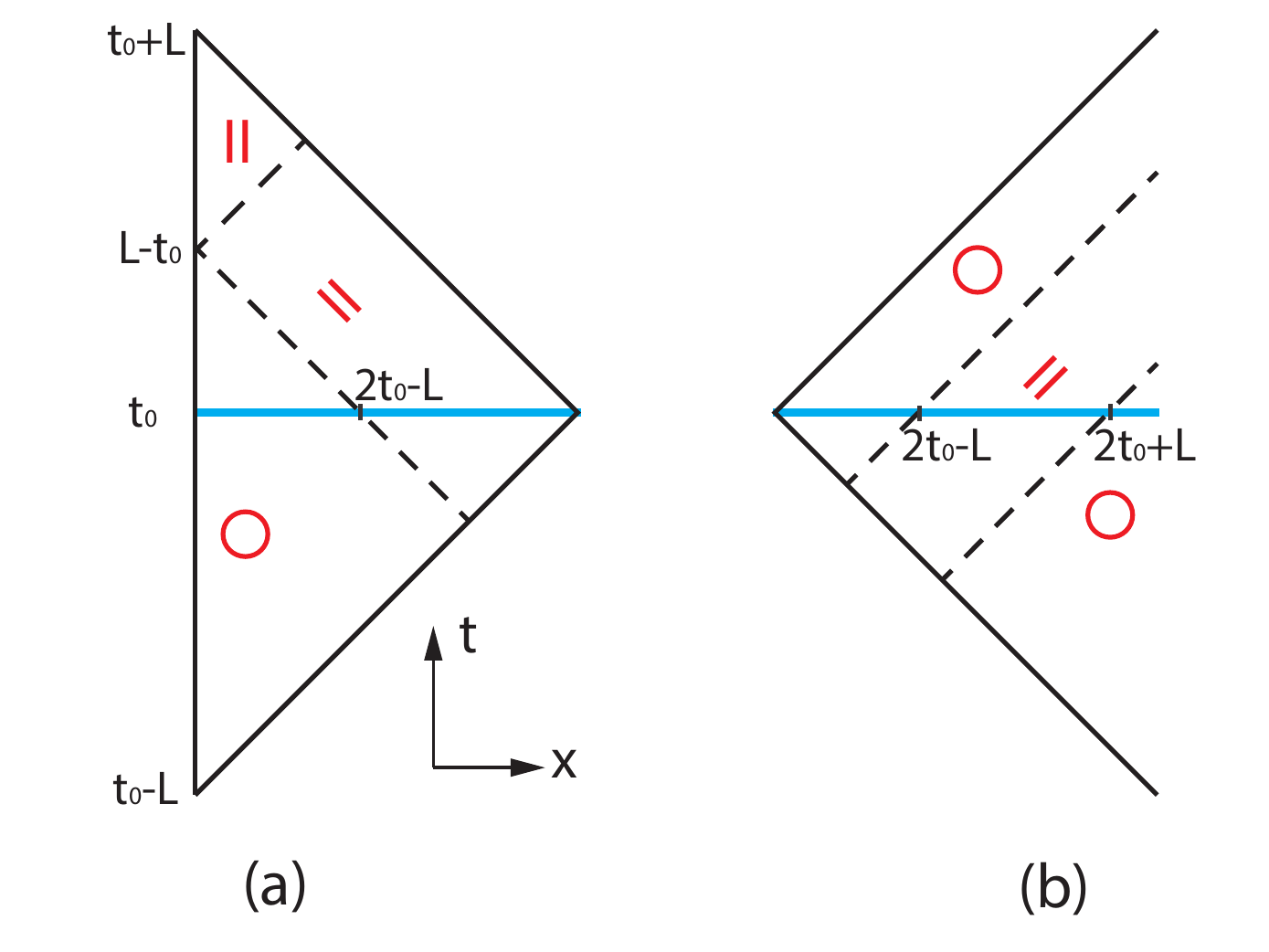}
\caption{(a) Causal wedge for subsystem $A=\{(x,t_0), 0\le x\le L\}$ (blue solid line) with $t_0<L$ in Minkowski spacetime,
with a physical boundary along $x=0$.
The wedge is divided into three regions labeled by
$|\,|$, $\backslash\backslash$, and $\bigcirc$, respectively,
as defined in Eqs.\ (\ref{RegionI_II}) and (\ref{Region_circle_I}).
(b) Causal wedge for subsystem $B=\{(x,t_0), L<x<\infty\}$ with $t_0>L$.
The wedge is
divided
 into  two regions $//$ and $\bigcirc$, as defined in
Eqs.\ (\ref{region//_B1}) and (\ref{region//_B2}).
}\label{wedge}
\end{center}
\end{figure}


\subsection{Flows in Minkowski spacetime for subsystem $[0,L]$}

Shown in Fig.$\,$$\,$\ref{wedge} (a) is the causal wedge for subsystem $A=[0,L]$ at $t_0$.
Here we denote by  $t_0$  the observation time, and by  $t$  the  Minkowski coordinate.
 To facilitate our later discussion, we divide the causal wedge into three regions labeled by symbols
 $|\,|$, $\backslash\backslash$, and $\bigcirc$ as follows:
\be\label{RegionI_II}
\text{region }|\,|:\quad
\left\{
\begin{split}
&t-(L-t_0)>x,\\
&t-(L+t_0)<-x,\\
&x>0,
\end{split}
\right.
\quad
\text{region }\backslash\backslash:\quad
\left\{
\begin{split}
&t-(L-t_0)<x,\\
&t-(L-t_0)>-x,\\
&t-(L+t_0)<-x,\\
&t-(t_0-L)>x,
\end{split}
\right.
\ee
and
\be\label{Region_circle_I}
\text{region }\bigcirc:\quad
\left\{
\begin{split}
&t-(L-t_0)<x,\\
&t-(L-t_0)<-x,\\
&t-(t_0-L)>x,\\
&t-(t_0-L)>-x.
\end{split}
\right.
\ee
It is noted that these regions are well defined for $t_0\le L$. For $t_0>L$, regions $\backslash\backslash$ and $\bigcirc$ will
shrink to zero, and region $|\,|$ will occupy the whole wedge.

Given the conformal mapping $w=f(z)$ in Eq.\ (\ref{two-step-half-line}), and  by considering $\text{Re} f(z)=u$ and making
the analytic
continuation $\tau\to it$, one 
obtains the equation describing these flows (compare also Eq.s (\ref{F_Eu01},\ref{F_Eu},\ref{Global_Homogeneous})
in Appendix B):
\begin{equation}\label{Flow_Eq}
\begin{split}
&\frac
{
\cosh\frac{2\pi}{\beta}\left(x-L\right)-\cosh\frac{2\pi}{\beta}\left(t-t_0\right)
}
{
\cosh\frac{2\pi}{\beta}\left(x+L\right)-\cosh\frac{2\pi}{\beta}\left(t-t_0\right)
}\cdot
\frac
{
\cosh\frac{2\pi}{\beta}\left(x+L\right)+\cosh\frac{2\pi}{\beta}\left(t+t_0\right)
}
{
\cosh\frac{2\pi}{\beta}\left(x-L\right)+\cosh\frac{2\pi}{\beta}\left(t+t_0\right)
}
=e^{-2u}.
\end{split}
\end{equation}

\begin{figure}
\begin{center}
\includegraphics[width=6.0in]{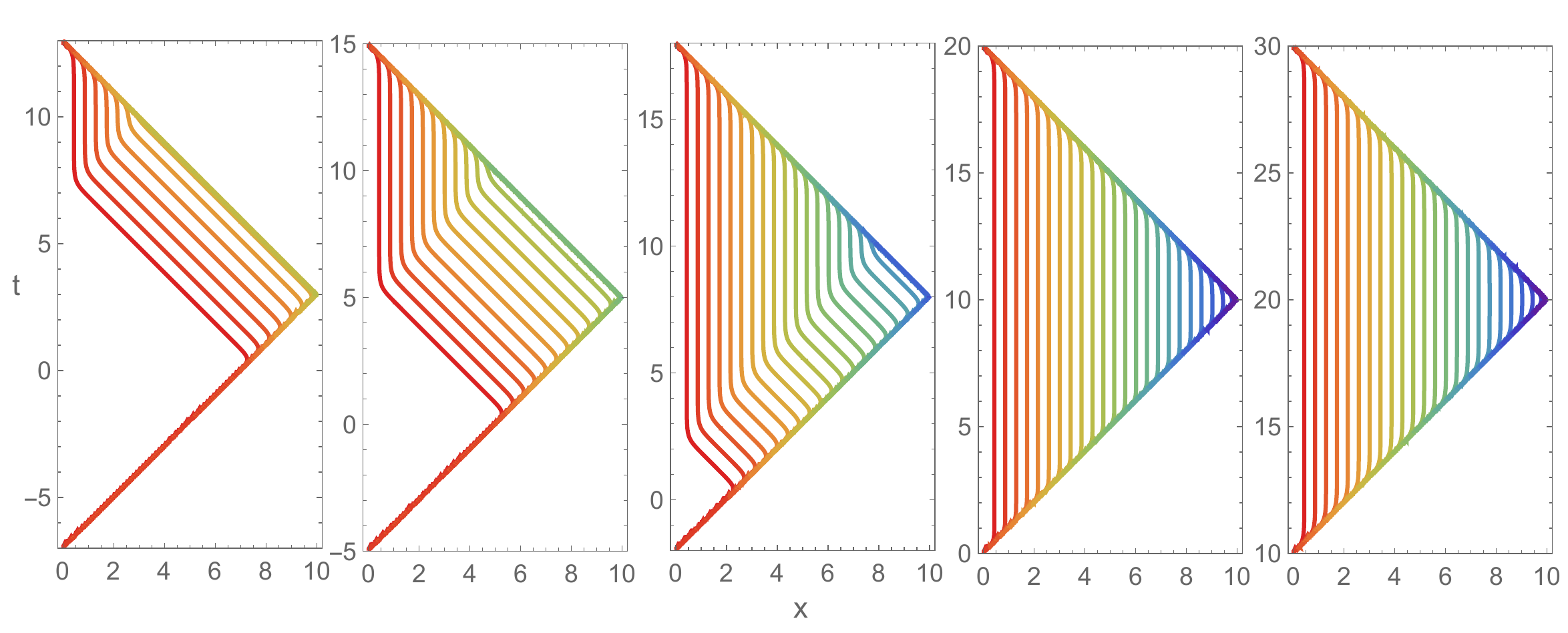}
\caption{Constant-$u$ flows in the causal wedge of subsystem $A=\{(x,t_0), 0\le x\le L\}$ in Minkowski spacetime,
plotted according to Eq.\ (\ref{Flow_Eq}).
The parameters we use are $\beta=1.5$, and $L=10$.
The observation times are $t_0=3,5,8,10,20$ from left to right.
Region $|\,|$ (see the definition in Fig.$\,$$\,$\ref{wedge}) is occupied by vertical flows,
region $\backslash\backslash$ is occupied by
left-tilted flows, and region $\bigcirc$ is occupied by nothing.
As time $t_0$ increases, the causal wedge is gradually occupied by region $|\,|$. For $t_0>L$,
the causal wedge is totally occupied by region $|\,|$ with vertical flows, which are independent of the observation time $t_0$.
See Eq.\ (\ref{Compare1}) for more quantitative interpretation.
}\label{FlowA}
\end{center}
\end{figure}

Based on
this
equation, 
we plot the constant-$u$ flows in Fig.$\,$$\,$\ref{FlowA}.
One
finds
that the result depends on the observation time $t_0$. 
We note the following
 interesting features:

(i) For $t_0<L$, one can always observe the  three different regions labeled by $|\,|$, $\backslash\backslash$, and $\bigcirc$
in Fig.$\,$$\,$\ref{wedge}. In particular, region $|\,|$ is filled with vertical flows, region $\backslash\backslash$ is filled with left-tilted
flows, and region $\bigcirc$ is empty (no flows).

(ii) As the observation time $t_0$ increases, regions $\backslash\backslash$ and $\bigcirc$
gradually shrink to zero, and the  region $|\,|$  of vertical flows gradually increases until $t_0=L$.
For $t_0\ge L$, the whole wedge is occupied by vertical flows, and the distribution of these
vertical flows 
is
 independent of $t_0$.
Upon
 comparing with Fig.$\,$$\,$\ref{ThermalState_Half}
for a finite interval $[0,L]$ at the end of
a semi-infinite system {\it at finite temperature} $\beta^{-1}$,
one observes
 that
the distributions of vertical flows are the same (See also Ref. \cite{ModularFlowThermal} for the modular flows
in a thermal ensemble.).
This indicates
that
 subsystem $A$ in Fig.$\,$$\,$\ref{FlowA} is thermalized for $t_0>L$.

Furthermore, as shown in Appendix \ref{ModuarFlow_Appendix},
 the flows in region $|\,|$ and region $\backslash\backslash$ can be approximately described by
\be\label{Vertical_Left}
\left\{
\begin{split}
&\text{flows in region }|\,|: \quad x=\frac{\beta}{2\pi}u,\\
&\text{flows in region }\backslash\backslash:\quad (x-L)+(t+t_0)=\frac{\beta}{\pi}u.
\end{split}
\right.
\ee
It is noted that the vertical flows described by $x=\beta u/2\pi$ are the
feature of a thermal ensemble \cite{ModularFlowThermal}.
 It agrees with Eq.\ (\ref{Vertical_FiniteT_Half}) for a thermal ensemble
at temperature $\beta^{-1}$, up to a global constant shift.
On the other hand, 
 left- (and right-) tilted constant-$u$ flows are
a feature of a global quench without
thermalization, as shown in Fig.$\,$\ref{GlobalFlowInfinite} in Appendix.
The second equation in Eq.\ (\ref{Vertical_Left})
agrees 
(up to a global
constant shift)
with Eq.\ (\ref{Left_tilted_B}) describing  
 a semi-infinite subsystem $A$ after a quantum quench.

Therefore, the time  evolution of the  constant-$u$ flows in Fig.$\,$$\,$\ref{FlowA} shows how
subsystem $A$ is
thermalized as $t_0$ increases.
Furthermore,  we can
obtain
a quantitative
correspondence between the flows in Fig.$\,$$\,$\ref{FlowA}
and the entanglement hamiltonian $K_A(t)$ in Eq.\ (\ref{KA_3_region}). By simply
looking at which kind of region intersects 
 subsystem $A=\{(x,t_0), 0\le x\le L\}$,
one finds
\begin{equation}\label{Compare1}
\begin{split}
(i)\,\,
t_0<L/2:\quad
\backslash\backslash \cap A&= [L-2t_0, L]
\longleftrightarrow
\int_{L-2t_0}^L\overline{T}(x,t_0)dx.\\
(ii)\,\,
L/2<t_0<L:\quad
|\,|\cap A&=[0, 2t_0-L]
\longleftrightarrow
\int_{0}^{2t_0-L} T_{00}(x,t_0)dx.\\
\backslash\backslash\cap A&=[2t_0-L, L]
\longleftrightarrow
\int_{2t_0-L}^L\overline{T}(x,t_0)dx.\\
(iii)\,\, t_0>L:\quad
|\,|\cap A&=[0, L]
\longleftrightarrow
\int_{0}^{L} T_{00}(x,t_0)dx.
\end{split}
\end{equation}
Based on this quantitative correspondence, we can conclude that the left-tilted flows in region $\backslash\backslash$
are contributed by the left-moving quasiparticles, and the vertical flows in region $|\,|$ are contributed by both
left-moving \textit{and}
right-moving quasiparticles.

\begin{figure}
\begin{center}
\includegraphics[width=6.0in]{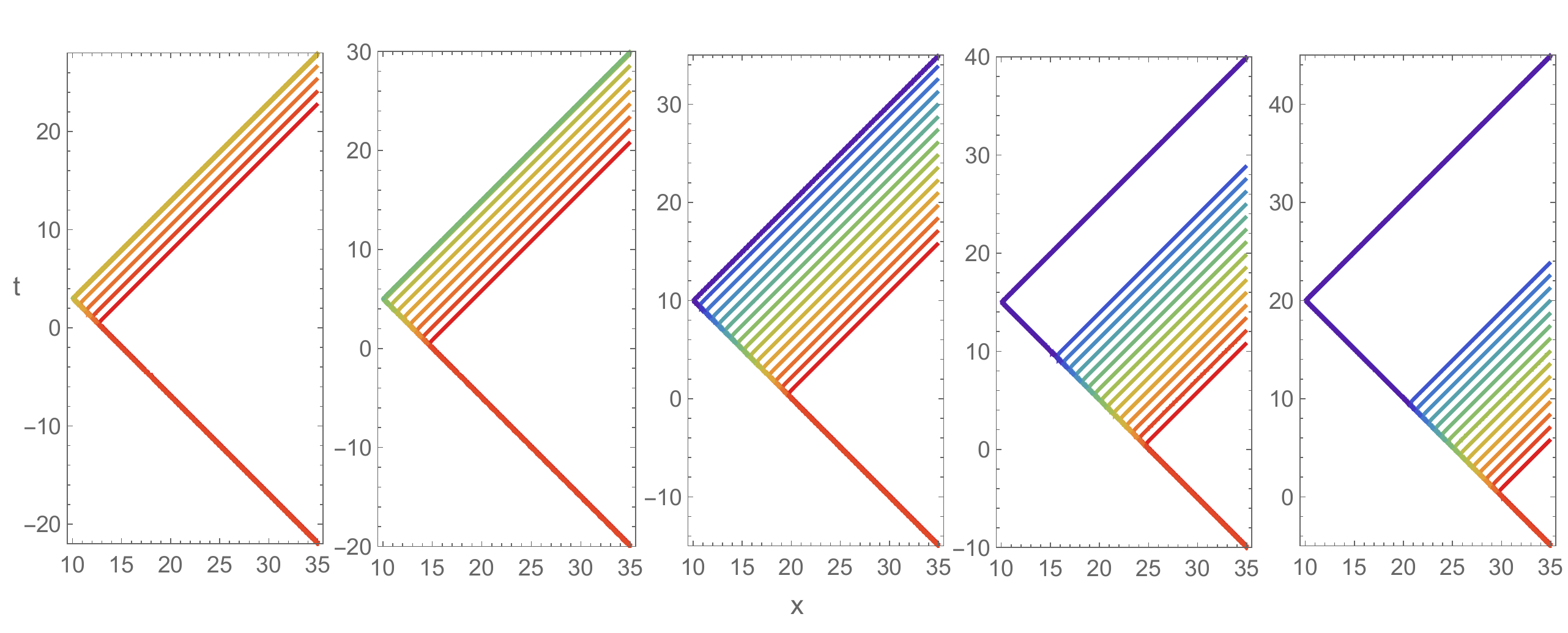}
\end{center}
\caption{
Constant-$u$ flows in the causal wedge of subsystem $B=\{(x,t_0), L<x<\infty\}$
in Minkowski spacetime, plotted according to Eq.\ (\ref{Flow_Eq}).
The parameters we use are $\beta=1.5$, and $L=10$.
From left to right, we have $t_0=3,5,10,15,20$.
For $t_0<L$, region $//$ (see the definition in Eqs.(\ref{region//_B1}) and
(\ref{region//_B2})), which is filled with right-tilted flows, grows as a function of $t_0$.
For $t_0>L$, region $//$ does not grow any more, but simply moves rightwards.
See Eq.\ (\ref{CompareB1})  for more interpretation.
}\label{FlowB}
\end{figure}

\subsection{Flows in Minkowski spacetime for subsystem $(L,\infty)$}

Now we consider the causal wedge for $B=\{(x,t_0), L<x<\infty\}$,
as shown in Fig.$\,$$\,$\ref{wedge} (b).
We 
divide  the wedge into two regions, region $//$ (which is filled with  right-tilted flows)
and the 
remaining  part (labeled by the symbol $\bigcirc$).
Region $//$ is defined as
\be\label{region//_B1}
\text{region }//:
\left\{
\begin{split}
&t-t_0<x-L,\\
&t-t_0>-(x-L),\\
&t-t_0>x-(2t_0+L),
\end{split}
\right.
\quad \text{for } t_0<L,
\ee
and
\be\label{region//_B2}
\text{region }//:
\left\{
\begin{split}
&t-t_0<x-(2t_0-L),\\
&t-t_0>-(x-L),\\
&t-t_0>x-(2t_0+L),
\end{split}
\right.
\quad \text{for } t_0>L.
\ee
The  corresponding constant-$u$ flows in Minkowski spacetime 
are
shown in Fig.$\,$$\,$\ref{FlowB}. There are several interesting features:

(i) For $t_0<L$, the region $//$ which is filled with right-tilted flows grows as a function of $t_0$.

(ii) For $t_0>L$, region $//$ does not grow any more as $t_0$ increases, but simply moves rightwards linearly in $t_0$.
For comparison, we study the constant-$u$ flows for the region  $A=[0,\infty]$ in an infinite system $(-\infty, +\infty)$
 after a global quench.
As shown Fig.$\,$$\,$\ref{GlobalFlowInfinite}, the region $//$ always grows as a function of time $t_0$, and never saturates.
This difference arises from the following facts: For the case in Fig.$\,$$\,$\ref{FlowB},
the number of quasiparticles carrying entanglement in $(L,\infty)$ will saturate due to the finite size of
its reservoir $[0,L]$.  For the case in Fig.$\,$$\,$\ref{GlobalFlowInfinite}, due to the semi-infinite
size 
 of both, of subsystem $A=[0,\infty)$
as well as of the reservoir
$B=(-\infty, 0)$, the number of quasiparticles carrying entanglement in $A$ will continue to grow  
as a function of
time $t_0$ without saturation. This agrees with the analysis
given in the paragraph below
 Eq.\ (\ref{KB_2_region}).

(iii) 
When compared with Fig.$\,$$\,$\ref{FlowA}, there are no vertical flows in Fig.$\,$$\,$\ref{FlowB}.
This is because there are only right-moving quasiparticles carrying entanglement in region $(L,\infty)$.

In addition, as shown in the appendix,
the flows in region $//$  can be approximately described by
\be\label{//_half_B}
\text{flows in region }//:\quad (x-L)-(t+t_0)=-\frac{\beta}{\pi}u.
\ee
Again, this is 
a feature of a global quantum quench without thermalization.
It agrees
(up to a global constant shift)
 with Eq.\ (\ref{Global_infinity_left_tilted}), which describes
 the right-tilted flows
for subsystem $A=[0,\infty)$ in an infinite system  $(-\infty, +\infty)$ after a global quantum quench.

Similarly, one can find the following quantitative correspondence between the constant-$u$ flows and
the entanglement hamiltonian,
\be\label{CompareB1}
\begin{split}
(i)\,\, t_0<L: \quad
// \cap B&=[L, 2t_0+L]
\longleftrightarrow
\int_{L}^{2t_0+L}T(x,t_0)dx.\\
(ii)\,\, t_0>L: \quad
// \cap B&=[2t_0-L, 2t_0+L]
\longleftrightarrow
\int_{2t_0-L}^{2t_0+L}T(x,t_0)dx,
\end{split}
\ee
where $T(x,t_0)$ is the
energy-momentum tensor for the right movers.
Based on the above analysis, one 
sees  that the right-tilted flows in region $//$ are contributed by the
right-moving quasiparticles.

In  short,
for the flows in Fig.$\,$$\,$\ref{FlowA} and Fig.$\,$$\,$\ref{FlowB}, we find that the left-tilted flows in region $\backslash\backslash$
and right-tilted flows in region $//$ are contributed by the left-moving and right-moving quasiparticles, respectively. The vertical flows
in region $|\,|$ are contributed by the left-moving 
{\it and}  the
 right-moving quasiparticles.
In region $\bigcirc$, there are no flows, and no quasiparticles in this region can contribute to entanglement.
The correspondence among the constant-$u$ flows,
quasiparticles (q.p.) carrying entanglement, and the entanglement hamiltonians can be summarized as
\be
\left\{
\begin{split}
\text{vertical flows in region }|\,| &\Leftrightarrow  \text{left}+ \text{right-moving q.p.} \Leftrightarrow  \int T_{00}(x,t)dx\\
\text{left-tilted flows in region }\backslash\backslash &\Leftrightarrow \text{left-moving q.p.} \Leftrightarrow  \int \overline{T}(x,t)dx\\
\text{right-tilted flows in region }//&\Leftrightarrow\text{right-moving q.p.} \Leftrightarrow  \int T(x,t)dx.\\
\end{split}
\right.
\ee
The time evolution of the entanglement hamiltonian and  the modular flows in Minkowski spacetime
provide us with  an intuitive picture on how the
entanglement propagates, and how the subsystem thermalizes.

\section{Discussion 
of  generic initial states}

\begin{figure}[htp]
\begin{center}
\includegraphics[width=5.00in]{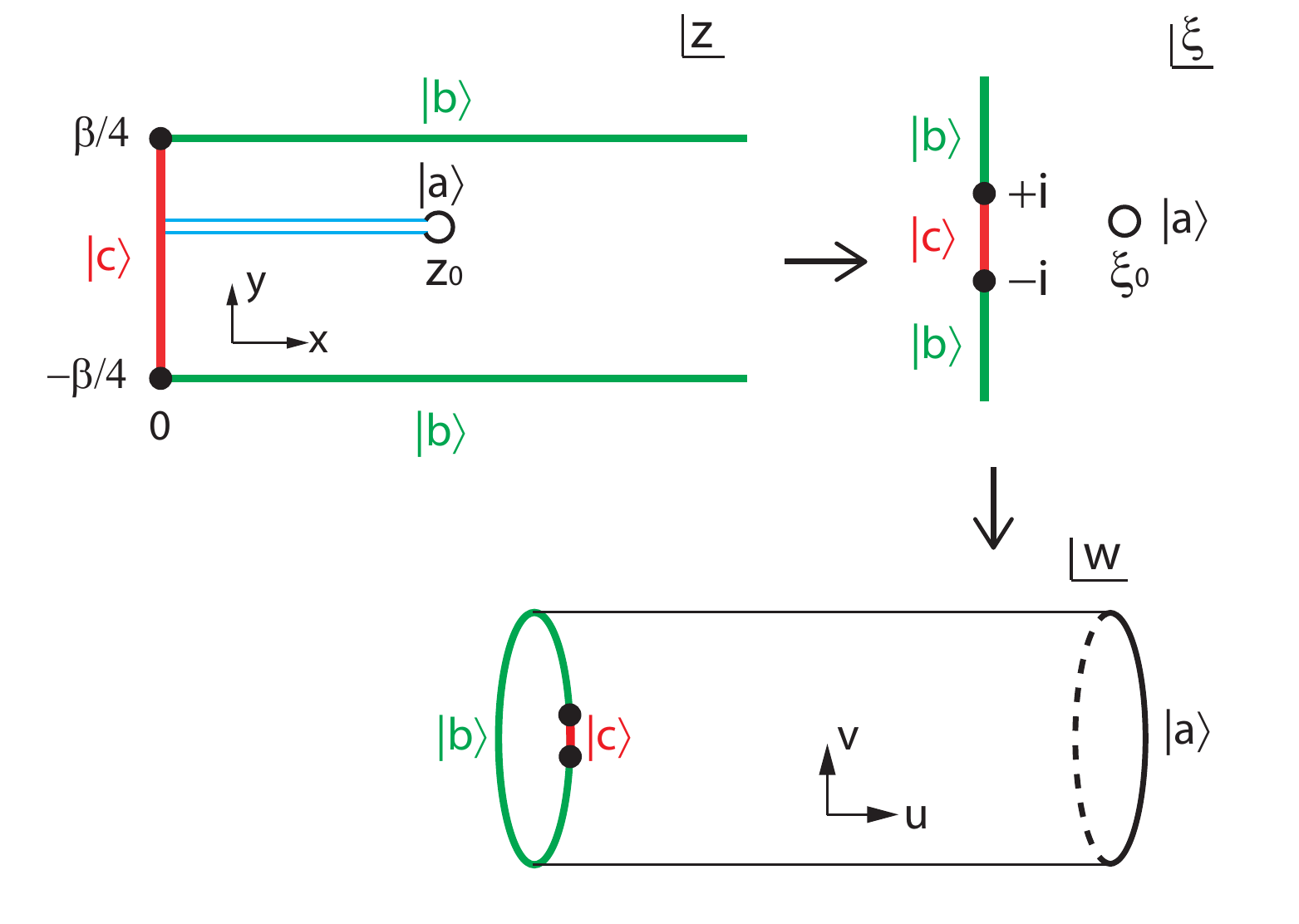}
\caption{Setup for a global quantum quench with $|b\rangle\neq |c\rangle$.
With the same conformal mapping in Eq.\ (\ref{two-step-half-line}), the half-rectangle
can be mapped to an annulus in the  $w$-plane. The black dots represent boundary condition
changing operators.
}
\label{ConformalMap_BoundaryChanging}
\end{center}
\end{figure}

Here, 
by  a generic {\it  initial} state, we mean  an initial state described by
$|\phi_0\rangle=e^{-(\beta/4)H_{\rm{CFT}}}|b\rangle$
corresponding to a regularized version of 
 a  conformal boundary state $|b\rangle$,
which
may be different from the conformal boundary state $|c\rangle$ describing  the {\it physical  boundary condition
at  position  $x=0$} (the left end of the semi-infinite region $[0,+\infty)$), as shown in
Fig.\ref{ConformalMap_BoundaryChanging}.
When $|b\rangle \not = |c\rangle$,
 the form of  the entanglement hamiltonian and of  the modular flows 
is
 the same as in
the case 
where  $|b\rangle=|c\rangle$.
However, one should be more careful about the boundary condition(s)  on the left edge of the annulus in the  $w$-plane
(see Fig.\ref{ConformalMap_BoundaryChanging}), which may affect the spectrum of the  entanglement hamiltonian.
In the following, we will analyze how this boundary condition evolves as a function of time after a quantum quench.

Let us focus on the boundary along 
$z \in \left \{ x\pm i\frac{\beta}{4}: x\ge 0 \right \}
 \cup \left\{ 0+i y: -\beta/4\le y \le \beta/4 \right\}$  
in the  $z$-plane
(Fig.\ref{ConformalMap_BoundaryChanging}):
The conformal boundary condition $|b\rangle$ is imposed along  the horizontal boundaries $z=\left(x\pm i\frac{\beta}{4}\right)$,
$x>0$, while 
the boundary condition $|c\rangle$
is imposed along the vertical boundary $z=\left(0+i y\right)$, $-\beta/4 < y < +\beta/4$.
With the two-step conformal mapping in Eq.\ (\ref{two-step-half-line}),
the entire
 boundary (consisting of horizontal and vertical pieces)  is mapped to the left edge of annulus in
the 
$w$-plane, which is a circle with circumference $2\pi$.
It is straightforward to check that this circle is along  $\text{Re}\,w=0$ in the  $w$-plane.
Now we are mainly interested in where the two boundary conditions  $|b\rangle$ and $|c\rangle$ are
located  along this circle.
Without loss of generality, let us study the
location
 of  the boundary condition $|b\rangle$  (boundary condition  $|c\rangle$ is
located in
the 
remaining interval of the circle, the complement), which is defined along
\be
w\left(x\pm i\frac{\beta}{4}\right)
=-\log
\left\{
\frac{
1+\sinh\left[\frac{2\pi}{\beta}(-i\tau+L)\right]
}
{
1+\sinh\left[\frac{2\pi}{\beta}(i\tau+L)\right]
}
\cdot
\frac{
\sinh\left[\frac{2\pi}{\beta}(x\pm i\frac{\beta}{4})\right]-\sinh\left[\frac{2\pi}{\beta}(i\tau+L)\right]
}{
\sinh\left[\frac{2\pi}{\beta}(x\pm i\frac{\beta}{4})\right]+\sinh\left[\frac{2\pi}{\beta}(-i\tau+L)\right]
}
\right\},
\ee
where $x\ge 0$. As shown in Fig.\ref{Edge_BoundaryChanging},
we use the following quantity to characterize how $|b\rangle$
wraps around the circle along $\text{Re}\,w=0$:
\be
\alpha(x):=w\left(x+i\frac{\beta}{4}\right)-w\left(x-i\frac{\beta}{4}\right).
\ee
After some algebra, and upon  making the analytical continuation $\tau\to it$, one 
obtains
\be\label{Alpha}
\alpha(x)=2\,\text{arctan}
\left(
\frac
{
4\cosh\frac{2\pi t}{\beta}\sinh\frac{2\pi L}{\beta}\cosh\frac{2\pi x}{\beta}
}
{
\cosh\frac{4\pi L}{\beta}-\cosh\frac{4\pi t}{\beta}-2\cosh^2\frac{2\pi x}{\beta}
}
\right).
\ee
The 
location  of the boundary condition  $|b\rangle$ as well as its effect on the  entanglement spectrum 
may be discussed in the following two time regimes:

\begin{figure}[htp]
\begin{center}
\includegraphics[width=4.50in]{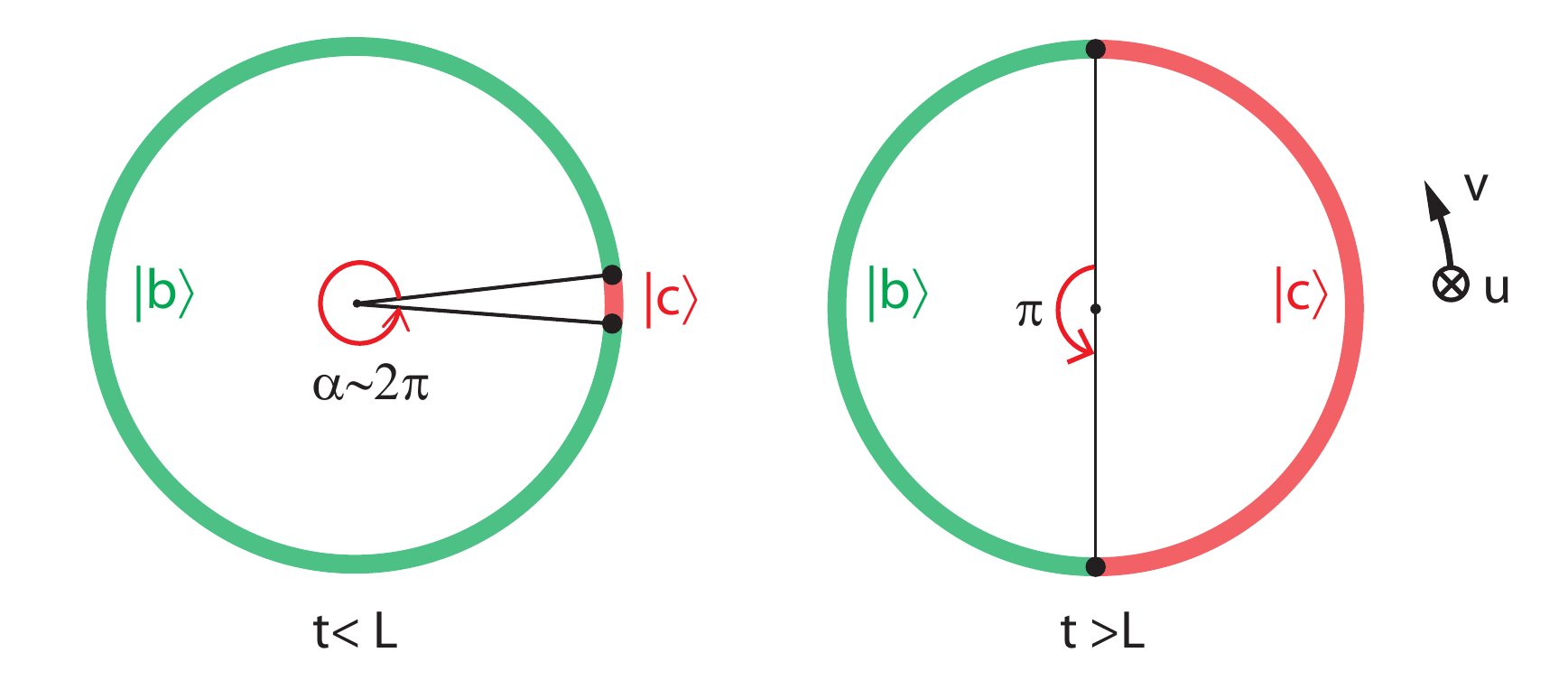}
\caption{Location
 of boundary conditions  $|b\rangle$ and $|c\rangle$ on the left edge of the annulus in the  $w$-plane, with $w=u+iv$
(see Fig.\ref{ConformalMap_BoundaryChanging}). For $t<L$, the edge is dominated by the conformal boundary
condition $|b\rangle$. For $t>L$, half the edge has boundary condition $|b\rangle$, and the other half
has boundary condition $|c\rangle$.
}
\label{Edge_BoundaryChanging}
\end{center}
\end{figure}

(i) $t<L$

Here, we are interested in the case $L-t\gg O(\beta)$.
In this time
regime,
 the entanglement entropy $S_A(t)$ grows linearly in time $t$ [see Eq.\ (\ref{SA(t)_I})].
Then $\alpha(x)$ in Eq.\ (\ref{Alpha}) can be simplified as
\be
\alpha(x)\simeq 2\,\text{arctan}\left(
\frac{
e^{\frac{2\pi}{\beta}(t+L+x)}
}{
e^{\frac{4\pi L}{\beta}}-e^{\frac{4\pi x}{\beta}}
}
\right)\in (0,2\pi),\quad \text{for}\,\, x\in [0,\infty).
\ee
That is, for $L-t\gg O(\beta)$, the boundary condition on the left edge of  the annulus in the  $w$-plane is dominated
by $|b\rangle$, and the effect of $|c\rangle$ can be neglected.
Then the entanglement spectrum is the same as that in the case $|c\rangle=|b\rangle$, as analyzed in
the previous section.
What is interesting is that, if one tracks back to the $z$-plane, one 
finds
 that the boundary condition $|b\rangle$
on the left edge of  the annulus is mainly contributed by the region
\be
x\in [L-t-O(\beta), L+t+O(\beta)],
\ee
which agrees with the quasi-particle picture in Fig.~\ref{EPR_Boundary}.
As shown in Fig.~\ref{EPR_Boundary} (a) and (b),
one sees that
the quasi-particles that contribute to
the entanglement entropy of subsystem $A$ are mainly emitted from the region $[L-t, L+t]$ in the initial state.

(ii) $t>L$

Now we are interested in the case $t-L\gg O(\beta)$.
In this time
regime,
 the entanglement entropy in subsystem $A$ saturates [see Eq.\ (\ref{SA(t)_I})].
Then $\alpha(x)$ in Eq.\ (\ref{Alpha}) may be simplified as
\be
\alpha(x)\simeq -2\,\text{arctan}\left(
\frac{
e^{\frac{2\pi}{\beta}(t+L+x)}
}{
e^{\frac{4\pi t}{\beta}}+e^{\frac{4\pi x}{\beta}}
}
\right)\in (\pi,2\pi),\quad \text{for}\,\, x\in [0,\infty).
\ee
That is, 
half of the circle has boundary condition $|b\rangle$, and the other half circle 
has boundary condition
$|c\rangle$,
as shown in the right panel of  Fig.\ref{Edge_BoundaryChanging}.
Again, if one tracks back to the  $z$-plane, the boundary condition $|b\rangle$ on the half-circle is mainly
contributed by the region
\be
x\in [t-L-O(\beta), t+L+O(\beta)].
\ee
This is consistent with the quasi-particle picture in Fig.\ref{EPR_Boundary} (c), from which
one sees
 that the entanglement between $A$ and $B$ is mainly contributed by the quasi-particles
emitted from the region $[t-L,t+L]$ in the initial state.

In this case, \textit{i.e.} when the boundary on the left edge of the cylinder in the $w$-plane
 is composed of both boundary conditions $|b\rangle$ and $|c\rangle$,
we do not know how to give an explicit form of entanglement spectrum.
(Though, technically, the presence of two different boundary conditions on the left edge
of the cylinder corresponds  to the presence of  two boundary condition changing operators\cite{CardyVerlindeNPB1989}.)
But if we study the entanglement entropy,
by repeating the calculations
in Eqs.\ (\ref{Tr_rhoA_N}) and (\ref{RvN_Entropy}), one 
finds
 that this specific boundary composed of two different boundary
conditions contributes only 
 a finite piece  to the
 entanglement entropy of order $\mathcal{O}(1)$, as compared to the previously discussed case
$|b\rangle =|c\rangle$ where a single boundary condition is imposed on the left edge of the cylinder.
The leading term of the  entanglement entropy is still given by $S_A\simeq \frac{c}{6}W$, where
$W$ is the width of annulus in Fig.\ref{ConformalMap_BoundaryChanging}
and its expression is given in Eq.\ (\ref{WA_I}).

One remark here. From the above analysis, one can conclude
 that even in the limit $t\to \infty$,
the information 
about the  initial state is still remembered by the finite subsystem $A$
(as opposed to what one would expect to find  for the thermalization process of a generic chaotic system, not
a conformal field theory):
Recall that in the global quench setup as discussed
in Ref. \cite{Calabrese2006},
certain 
correlation functions,  \textit{e.g.}, the 
one-point correlation function, always remember
the information of the initial state. (Its  amplitude is a matrix element of the conformal boundary state $|b\rangle$
characterizing the initial state.)
 Our conclusion agrees with this observation.

As a short summary, in the case $|b\rangle=|c\rangle$, the reduced density matrix $\rho_A(t\to\infty)$
is exactly the same as that
at
 a finite temperature $\beta$. But for $|b\rangle\neq |c\rangle$,
because the left boundary condition of the annulus is composed of both $|b\rangle$ and $|c\rangle$ segments,
the limit
$\rho_A(t\to \infty)$
 of the reduced density matrix  does not match
exactly  the thermal density matrix with external boundary
condition $|c\rangle$ (recall that $|c\rangle$  describes the boundary condition at the physical boundary $x=0$). The latter
density matrix  is illustrated  in  Fig.\ref{half_thermal}.
- I.e., in general, the limit 
 $\rho_A(t\to \infty)$
of the  reduced density matrix  still retains memory of the initial state  $|b\rangle$  (compare Fig. \ref{ConformalMap_BoundaryChanging}).

\section{Concluding remarks}

In this work, we study the time evolution of  the entanglement hamiltonian and related quantities
for a finite interval of length $L$ at the end of a semi-infinite system after a global quantum quench
into a (1+1) dimensional CFT
from a special class of {\it  initial}  states,
which are chosen to be  the same conformal boundary states as those describing the {\it physical boundary}  at the end $x=0$ of
semi-infinite space.
The results can be briefly summarized as follows.

-- For times  $t<L$, when the subsystem is not thermalized,
the entanglement hamiltonian depends on both the hamiltonian density and the momentum density.
After time $t=L$, when the subsystem $A$ is thermalized, the entanglement hamiltonian
only depends on the hamiltonian density.
In the long time limit $t\to \infty$, the entanglement hamiltonian (and therefore the reduced density matrix)
for subsystem $A$ is exactly the same as that in a thermal ensemble at finite temperature $\beta^{-1}$.

-- Using
conformal mappings and the knowledge of boundary CFT,
one can obtain
both the entanglement entropy and the entanglement spectrum
at arbitrary time $t$.
In particular, for $t>L$, it is found that the spacing of
entanglement spectrum approaches its long-time saturation value, \textit{i.e.} that of the entanglement spectrum
in a thermal ensemble at temperature $\beta^{-1}$, exponentially in time.

-- The modular flows in the causal wedge of subsystem $A$ in Minkowski spacetime are studied.
These flows provide us  with very rich information on how the subsystem is thermalized.
For $t<L$, these flows show a mixed feature of a thermal ensemble and a global quantum quench without thermalization.
As time evolves, the feature of a thermal ensemble dominates gradually.
For $t>L$, the distribution of the modular flows is independent of time, and looks the same as 
that of
 a thermal ensemble at finite temperature $\beta^{-1}$.
In addition, we find a quantitative correspondence between these flows and the
corresponding entanglement hamiltonians,
as shown in Eq.\ (\ref{Compare1}).
There are also interesting features 
of
 the modular flows corresponding to subsystem $B=(L,\infty)$,  where thermalization never occurs, which we discuss.

-- We also studied the case where the conformal boundary state  describing the
{\it  initial}  state is different from
that 
describing the {\it physical boundary}  at the end
 $x=0$ of semi-infinite space. It is found that even in the limit $t\to \infty$, the reduced density
matrix $\rho_A$ is not exactly the
same as that in a thermal state.
This is as expected for a rational CFT which is integrable and  where the full set  of conformal boundary states
is
 well
characterized.
Curious readers may ask what happens for an irrational CFT with a large central charge $c$.
In the later case, however,  it appears that we do not control the  set of
conformal boundary states (which should be composed of
an infinite number of Ishibashi states in irrational CFTs with a discrete set of primaries) 
and of
 corresponding boundary condition changing operators.
We note that our method  could presumably be extended 
 to irrational CFTs with a discrete spectrum of primaries,  once the corresponding properties of the conformal boundary
states and boundary condition changing operators are suitably 
 well understood.

\begin{figure}
\begin{center}
\includegraphics[width=5.5in]{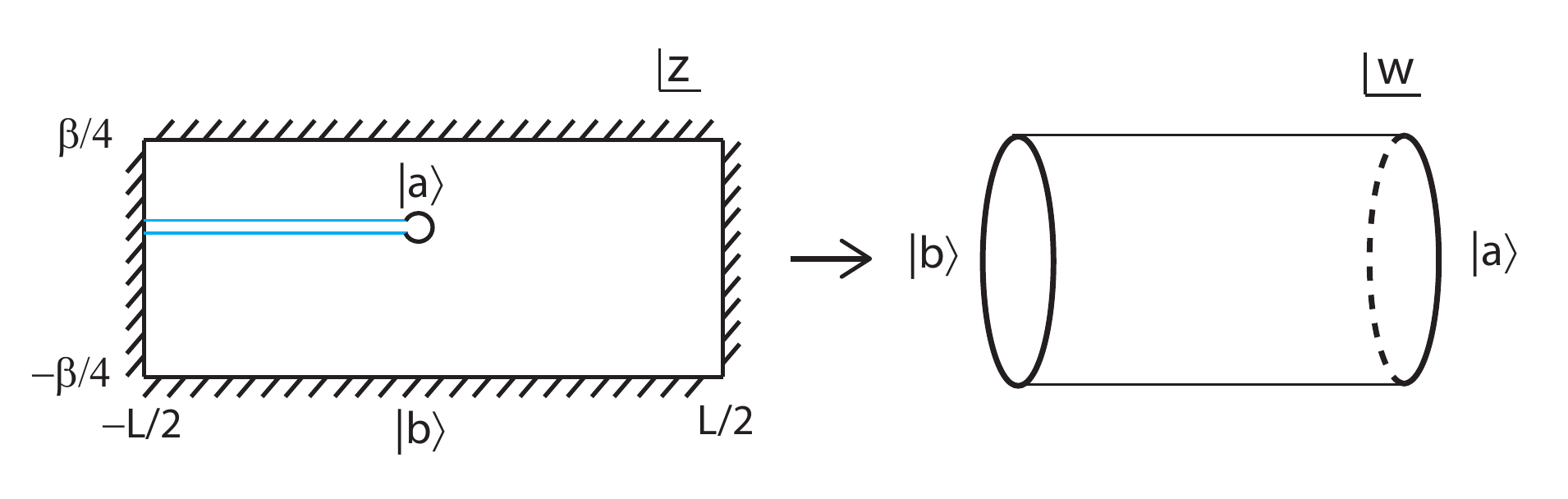}
\end{center}
\caption{
Setup for a subsystem $A=[-L/2,R]$ in a
finite system $[-L/2,L/2]$ after a global quantum quench. The height of the rectangle is $\beta/2$.
A small disc is removed at the entangling point as regularization, and the solid blue lines are branch cuts.
The topology of a rectangle with a small disc removed is equivalent to an annulus.
Therefore, one can use a conformal mapping $w=f(z)$ to map the rectangle in the  $z$-plane to an annulus in
the $w$-plane.
}\label{Finite_Quench}
\end{figure}

There are some future problems to study in detail:

-- It would be  interesting to study the time evolution of the entanglement hamiltonian and  the corresponding modular flows for an interval in
a \textit{finite} system after a global quantum quench. It is known that quantum revival may be observed
for a rational CFT due to the compact
nature  of the system \cite{Cardy1403}.
It is expected that
 revival of the entanglement hamiltonian and of the modular flows should also
be observed here.
The setup for studying a finite system after a global quantum quench is shown in Fig.$\,$$\,$\ref{Finite_Quench},
where we have a rectangle in the  $z$-plane with $z=x+iy$, $x\in [-L/2,L/2]$ and $y\in[-\beta/4,\beta/4]$.
For simplicity, one can impose again (analogous to what was done  in the main text of the present paper)
 the same conformal boundary condition along  the horizontal boundary $x=\pm L/2$ as on the vertical boundary
$y = \pm \beta/4$.
Then, upon choosing the subregion $A$ to be a finite interval {\it at the end} of the finite position space $[-L/2, +L/2]$,
 the topology of this
rectangle with a small disc removed  at the entangling point is topologically equivalent to an annulus, 
as shown in Fig.$\,$$\,$\ref{Finite_Quench}.
We can map the rectangle in the  $z$-plane to an annulus in the  $w$-plane based on a two-step conformal mapping:
One can first map the rectangle in the complex $z$-plane to  the right half complex plane (RHP)
$\xi$-plane by using
the
Schwarz-Christoffel  transformation (see e.g. Ref. \cite{Mandal1604}). Then the RHP with a small disc at $\xi_0$ removed
can be mapped to an annulus in the $w$-plane by using the second formula in Eq.\ (\ref{two-step-half-line}).

-- It would be interesting to also  study the case of inhomogeneous quantum quenches. In the current work,
since the global quantum quench evolves from an initial state that is translation invariant,
the density of modular flows is homogeneous and proportional to $\beta^{-1}$.
For certain inhomogeneous quantum quenches, the inverse temperature is
 a function
$\beta(x)$
 of spatial position $x$, which indicates that
the correlation length of the initial state is position-dependent.
It is expected that one can observe modular flows with inhomogeneous density.
Note that in Ref.\ \cite{Cardy1608}, some general results and features of 
the evolution of the entanglement hamiltonian  in inhomogeneous quenches have been studied,
though  it is still interesting to study certain concrete setups of inhomogeneous quantum quenches. 
See,
\textit{e.g.}, the setups proposed in Refs.\ \cite{InhomogeneousQuench}.
In particular, inhomogeneous hamiltonians are introduced in some setups, which are beyond the cases studied in [27].
 (We note that the entanglement hamiltonian
for certain inhomogeneous (1+1)d CFTs in the ground state has been studied most recently \cite{Tonni2017}.)

\section{Acknowledgements}

We are grateful to the KITP Program “Quantum Physics of Information” (Sep 18 - Dec 15, 2017).
This work was supported by
the Gordon and Betty Moore Foundation's EPiQS initiative through Grant No. GBMF4303 at MIT (XW),
the NSF under Grants No. NSF PHY-1125915 (SR),
and the NSF under Grant No. DMR-1309667 (AWWL).

\begin{appendices}

\section*{Appendices}

\section{On conformal mappings, etc}
\label{app: maps}

\subsection{On entanglement hamiltonians}

\subsection*{Entanglement hamiltonians for subsystem $A$ at the end of a semi-infinite system}
\label{Subsection: EH}

For Eq.\ (\ref{KAtotal_half}), first let us check the region $\beta\ll t<L$.
By considering the limit $t,L\gg \beta$ and ignoring the contribution near
$L-x\sim O(\beta)$, one can make the approximation $\sinh\left[\frac{\pi}{\beta}(x-L)\right]\simeq -\frac{1}{2}e^{\frac{\pi}{\beta}(L-x)}$.
Then Eq.\ (\ref{KAtotal_half}) can be approximated as
\begin{equation}\label{KAI_i}
\begin{split}
K_A(t<L)
\simeq&
\frac{\beta}{\pi}\int_{0}^{L}\frac{1}{2}\cdot
\frac
{
e^{\frac{\pi}{\beta}(L-x)}\times \cosh\left[\frac{\pi}{\beta}(x-2t+L)\right]
}
{
\cosh\left[
\frac{2\pi}{\beta}(x-t)
\right]
}T(x-t)dx\\
&+
\frac{\beta}{\pi}\int_0^L
\frac
{
1+
 e^{-\frac{2\pi}{\beta}(x+2t-L)}
}
{
2
}\overline{T}(x+t)dx.\\
\end{split}
\end{equation}
One 
finds
 that the result depends on whether $t<L/2$ or $L/2<t<L$ as follows.
For $t<L/2$, one has
\begin{equation}
K_A\left(t<L/2\right)\simeq \frac{\beta}{2\pi}\int_{L-2t}^L
\overline{T}(x+t)dx.
\end{equation}
For ${L}/{2}<t<L$, one can check that
\begin{equation}
\begin{split}
K_A\left(L/2<t<L\right)\simeq&
\frac{\beta}{2\pi}\int_0^{2t-L}T(x-t)dx
+\frac{\beta}{2\pi}\int_0^L\overline{T}(x+t)dx\\
=&\frac{\beta}{2\pi}\int_0^{2t-L}T_{00}(x,t)dx+\frac{\beta}{2\pi}\int_{2t-L}^L\overline{T}(x+t)dx.
\end{split}
\end{equation}
Now let us check the $t>L$ case. Again, by  ignoring the contribution near
$L-x\sim O(\beta)$, so that $\sinh\left[\frac{\pi}{\beta}(x-L)\right]\simeq -\frac{1}{2}e^{\frac{\pi}{\beta}(L-x)}$,
and considering the limit $t,L\gg \beta$,
$K_A(t)$ can be approximated as
\begin{equation}\label{KAI_ii}
\begin{split}
K_A(t>L)
\simeq&
\frac{\beta}{\pi}\int_L^0\frac{1}{2}\cdot
\frac{
-e^{\frac{\pi}{\beta}(L-x)}\times e^{\frac{\pi}{\beta}(2t-x-L)}\times
e^{\frac{\pi}{\beta}(x+L)}\times e^{\frac{\pi}{\beta}(2t+L-x)}
}
{e^{\frac{2\pi}{\beta}t}\times e^{\frac{2\pi}{\beta}L}\times e^{\frac{2\pi}{\beta}(t-x)}
}T(x-t)dx\\
&+
\frac{\beta}{\pi}\int_L^0\frac{1}{2}\cdot
\frac
{
-e^{\frac{\pi}{\beta}(L-x)}\times e^{\frac{\pi}{\beta}(x+2t+L)}\times e^{\frac{\pi}{\beta}(x+L)}\times e^{\frac{\pi}{\beta}(x+2t-L)}
}
{
e^{\frac{2\pi}{\beta}t}\times e^{\frac{2\pi}{\beta}L}\times e^{\frac{2\pi}{\beta}(x+t)}
}\overline{T}(x+t)dx\\
=&\frac{\beta}{2\pi}\int_0^L \left[T(x-t)+\overline{T}(x+t)\right]dx\\
=&\frac{\beta}{2\pi}\int_0^L  T_{00}(x,t)dx.
\end{split}
\end{equation}
In other words, $K_A(t)$ is proportional to the physical hamiltonian $H_A=\int_0^L  T_{00}(x,t)dx $.

\subsection{On modular flows in Minkowski spacetime}
\label{ModuarFlow_Appendix}

In this part, we give details of  the behavior of 
the
constant-$u$ flows in Fig.$\,$$\,$\ref{FlowA} and Fig.$\,$$\,$\ref{FlowB}.

\subsection*{Vertical flows in region $|\,|$ in Fig.$\,$$\,$\ref{FlowA}}

Let us check the case $t_0<L$ first. Then region $|\,|$ in Fig.$\,$$\,$\ref{FlowA}
is defined by the first formula in Eq.\ (\ref{RegionI_II}), based on which one can obtain
\be\label{Vertical_Diamond_01}
L+t_0>t+x>t-x>L-t_0>0.
\ee
By also considering $t_0<L$, one
finds
\be
|t-t_0|<L-x, \quad 0<L+x<t+t_0.
\ee
Then Eq.\ (\ref{Flow_Eq}) can be approximated as
\be\label{Vertical01}
\frac{e^{\frac{2\pi}{\beta}(L-x)}}{e^{\frac{2\pi}{\beta}(L+x)}}
\cdot
\frac{e^{\frac{2\pi}{\beta}(t+t_0)}}{e^{\frac{2\pi}{\beta}(t+t_0)}}=e^{-2u}
\quad
\Rightarrow \quad x=\frac{\beta}{2\pi}u.
\ee

Now let us consider the case $t_0>L$.  Then region $|\,|$ occupies the whole wedge, which is defined by
\be
|t-t_0|<L-x<L+x, \quad x>0.
\ee
In addition, by considering $t_0>L$, one can obtain $t+t_0>L+x$.
Then Eq.\ (\ref{Flow_Eq}) can be approximated as
\be\label{Vertical02}
\frac{e^{\frac{2\pi}{\beta}(L-x)}}{e^{\frac{2\pi}{\beta}(L+x)}}
\cdot
\frac{e^{\frac{2\pi}{\beta}(t+t_0)}}{e^{\frac{2\pi}{\beta}(t+t_0)}}=e^{-2u}
\quad
\Rightarrow \quad x=\frac{\beta}{2\pi}u.
\ee

\subsection*{Left tilted flows in region $\backslash\backslash$  in Fig.$\,$$\,$\ref{FlowA}}

As shown in Fig.$\,$$\,$\ref{FlowA}, the region $\backslash\backslash$ filled with  left-tilted flows is defined by
the second formula in Eq.\ (\ref{RegionI_II}), based on which one can find
\be
|t-t_0|<L-x<t+t_0<L+x.
\ee
Then Eq.\ (\ref{Flow_Eq}) can be approximated as
\be
\frac{
e^{\frac{2\pi}{\beta}(L-x)}
}{
e^{\frac{2\pi}{\beta}(L+x)}
}
\cdot
\frac{
e^{\frac{2\pi}{\beta}(L+x)}
}{
e^{\frac{2\pi}{\beta}(t+t_0)}
}=e^{-2u}
\quad
\Rightarrow
\quad
x-L+(t+t_0)=\frac{\beta}{\pi}u.
\ee

\subsection*{Right tilted flows in region $//$ in Fig.$\,$$\,$\ref{FlowB}}

Let us check the case $t_0<L$ first. The region $//$ is defined in Eq.\ (\ref{region//_B1}), based on which we 
find $|t-t_0|<x-L<t+t_0<x+L$. Then Eq.\ (\ref{Flow_Eq}) can be simplified as
\be
e^{\frac{2\pi}{\beta}(x-L-t-t_0)}=e^{-2u} \quad \Rightarrow\quad x-L-(t+t_0)=-\frac{\beta}{\pi}u.
\ee
Now let us consider the case $t_0>L$.
Region $//$ is defined in Eq.\ (\ref{region//_B2}), based on which we still have
$|t-t_0|<x-L<t+t_0<x+L$. Then, again, Eq.\ (\ref{Flow_Eq}) can be simplified as
\be
\quad x-L-(t+t_0)=-\frac{\beta}{\pi}u.
\ee

\begin{figure}
\begin{center}
\includegraphics[width=5.5in]{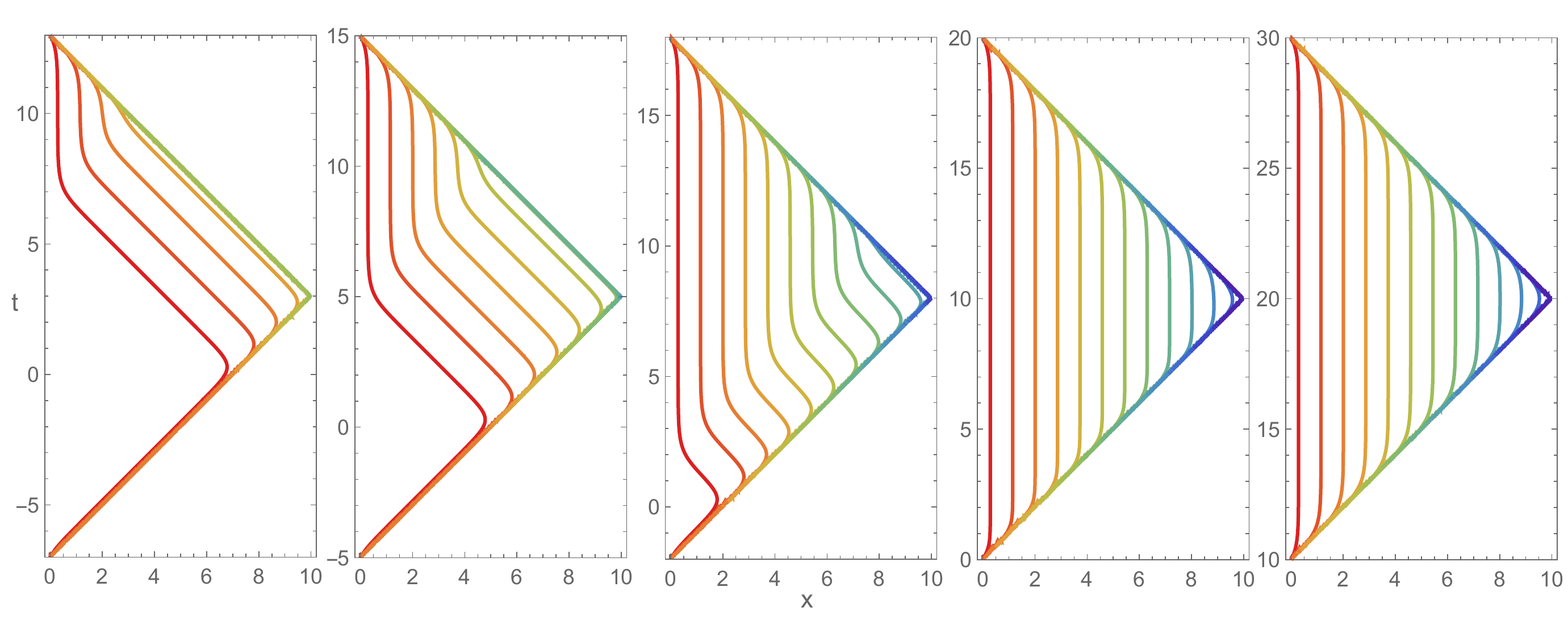}
\end{center}
\caption{
Constant-$u$ flows in the causal wedge of subsystem $A=\{(x,t_0), 0\le x\le L\}$ at the end of
a semi-infinite system in Minkowski spacetime.
The parameters we use are $\beta=3.0$, and $L=10$.
The observation times are chosen as
$t_0=3,5,8,10,20$ from left to right.
}\label{Diamond_T2}
\end{figure}

\subsection{Effect of $\beta$ on modular flows for a finite system after a global quench}

Eqs. (\ref{Vertical_Left}) and (\ref{//_half_B}) that describe the flows in Fig.$\,$$\,$\ref{FlowA} and Fig.$\,$$\,$\ref{FlowB}
are obtained in the limit $L\pm x, |t-t_0|\gg \beta$. It is natural to ask what happens
if we increase $\beta$ in the initial state $e^{-(\beta/4)H_{\rm{CFT}}}|b\rangle$?
Here we take the case in Fig.$\,$$\,$\ref{FlowA} for example.
As we increase $\beta$, one
finds
that the constant-$u$ flows
near the boundaries of regions $|\,|$ and $\backslash\backslash$
are no longer well approximated by straight lines, as shown in Fig.$\,$$\,$\ref{Diamond_T2}.
In fact, as we further increase $\beta$,
so that $\beta\ge L$, there will be no straight lines in the causal wedge.
This can be easily understood by considering the limit $\beta\to \infty$,
which corresponds to a CFT in its ground state, and there is essentially no quantum quench.

\section{Modular flows in Minkowski spacetime for a thermal ensemble, etc}

\subsection{Interval at the end of a semi-infinite system at finite temperature}

\begin{figure}[htp]
\begin{center}
\includegraphics[width=5.8in]{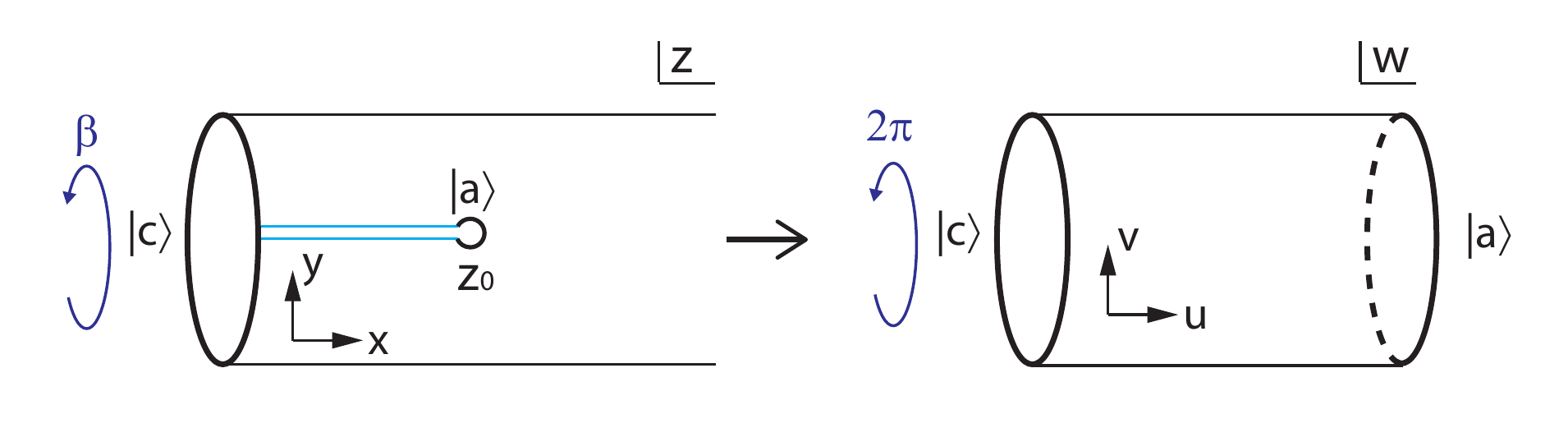}
\end{center}
\caption{
 Euclidean spacetime for $\rho_A$ at finite temperature $\beta$, with $A=[0,L]$.
}\label{half_thermal}
\end{figure}

For a finite interval $A=[0,L]$ at the end of a semi-infinite system at finite temperature $\beta^{-1}$,
we have a semi-infinite annulus of circumference $\beta$ in the imaginary time $\text{Im} (z)$ direction.
Again, we remove a small disc around the entangling point $L+i\tau$, where we can simply choose $\tau=0$.
In addition, we impose boundary conditions described by
 conformal boundary states $|a\rangle$ and $|c\rangle$ at the small disc and along the physical boundary
$x=0$, respectively. Then, by using the following conformal mapping
\begin{equation}\label{Map_finiteT}
w=f(z)=\log \frac{e^{\frac{2\pi z}{\beta}}-e^{-\frac{2\pi L}{\beta}}}{e^{\frac{2\pi z}{\beta}}-e^{\frac{2\pi L}{\beta}}},
\end{equation}
we map the semi-infinite annulus to a finite annulus in the $w$-plane, with conformal boundary conditions $|a\rangle$ and $|c\rangle$
at the two edges of the annulus. The circumference of the annulus in $\text{Im} \,w$ direction is $2\pi$.
The width of the annulus in the $w$-plane can be obtained by
\be
W=f(L-\epsilon)-f(0)=\log\left(
\frac
{
e^{\frac{2\pi}{\beta}(L-\epsilon)}-e^{-\frac{2\pi}{\beta}L}
}{
e^{\frac{2\pi}{\beta}(L-\epsilon)}-e^{\frac{2\pi}{\beta}L}
}
\cdot
\frac
{
1-e^{\frac{2\pi}{\beta}L}
}
{
1-e^{-\frac{2\pi}{\beta}L}
}
\right),
\ee
which can be rewritten as
\be\label{Thermal_half_W}
W= \log\frac{
\sinh[\pi(2L-\epsilon)/\beta]
}
{\sinh(\pi \epsilon/\beta)}.
\ee
In the limit $L\gg \beta\gg \epsilon$, $W$ can be further simplified as
\be\label{Thermal_W}
W\simeq\frac{2\pi}{\beta}L+\log\frac{\beta}{2\pi\epsilon}.
\ee
Based on the definition in Eq.\ (\ref{RvN_Entropy}),
the von Neumann entropy has the form (keeping only  the leading term in $L$)
\be\label{Thermal_vN_Entropy}
S_A\simeq \frac{\pi c}{3\beta} \cdot L.
\ee
In addition, it is straightforward to find the exact form of entanglement hamiltonian as follows
\be\label{FiniteT_KA_semi_infinite}
K_A=\frac{\beta}{\pi}\int_{0}^L\frac
{\sinh[\pi(L-x)/\beta]\sinh[\pi(L+x)/\beta]}
{\sinh(2\pi L/\beta)}T_{00}(x)dx,
\ee
which may be simplified as
\be
K_A(\beta)\simeq \frac{\beta}{2\pi}\int_{0}^{L}T_{00}(x)dx,
\ee
by ignoring the contributions near the entangling points $|L\pm x|\sim O(\beta)$.

\begin{figure}
\begin{center}
\includegraphics[width=3.2in]{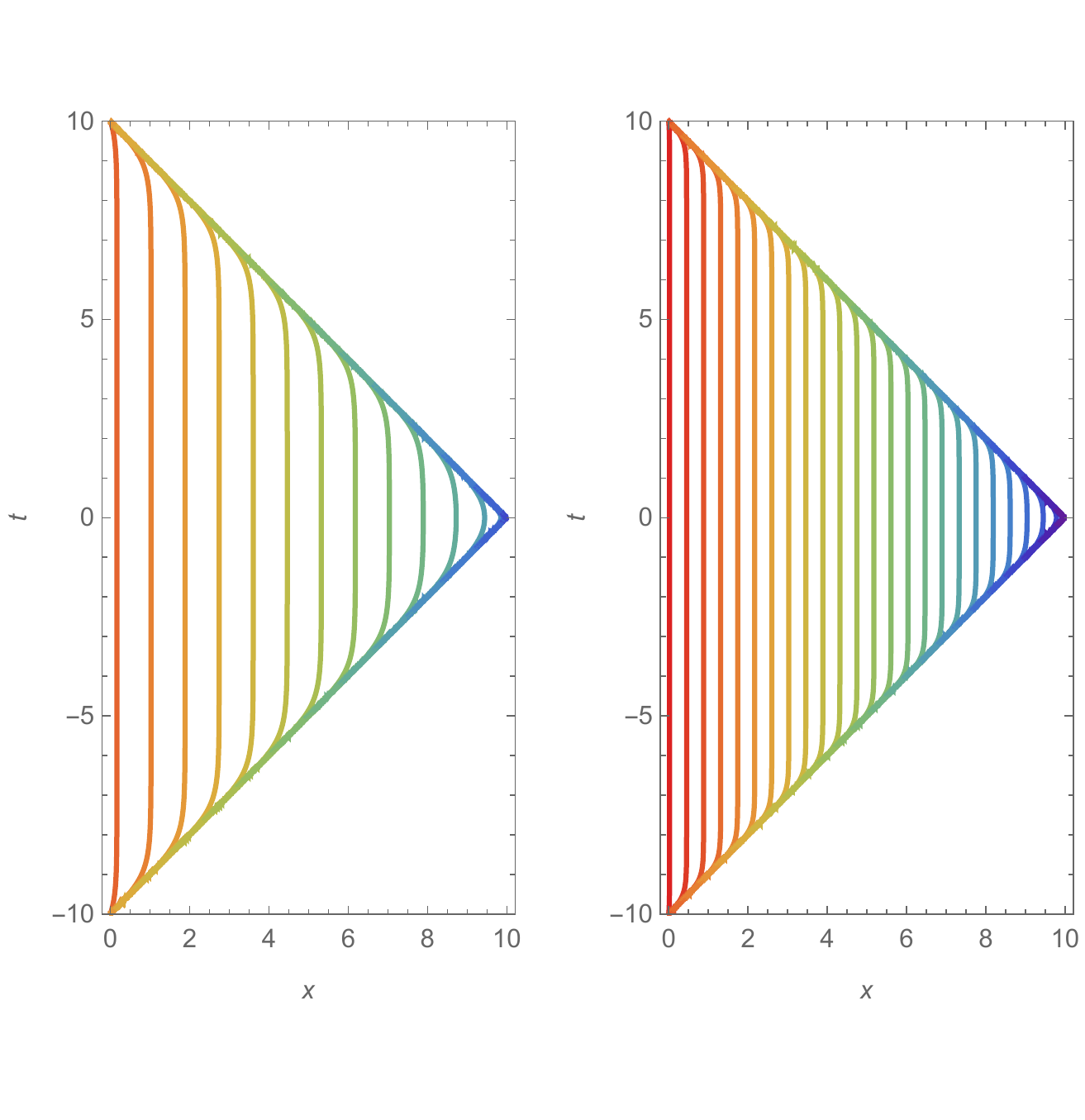}
\end{center}
\caption{
Constant-$u$ flows in the causal wedge of subsystem $A=\{(x,0), 0\le x\le L\}$ at finite temperature $\beta^{-1}$
in Minkowski spacetime. The physical boundary is along $x=0$.
The parameters we use are $L=10$, $\beta=3.0$ (left) and $\beta=1.5$ (right).
}\label{ThermalState_Half}
\end{figure}

Now, we will study the constant-$u$ flows for subsystem $A$ in Minkowski spacetime.
Based on the conformal mapping in Eq.\ (\ref{Map_finiteT}), one can get
\be
\frac
{
e^{\frac{4\pi}{\beta}x}-2e^{\frac{2\pi}{\beta}(x-L)}\cos\frac{2\pi}{\beta}y+e^{-\frac{2\pi}{\beta}2L}
}
{
e^{\frac{4\pi}{\beta}x}-2e^{\frac{2\pi}{\beta}(x+L)}\cos\frac{2\pi}{\beta}y+e^{\frac{2\pi}{\beta}2L}
}
=e^{2u}.
\ee
Making the  analytic continuation $y\to it$, one can further obtain the flows in Minkowski spacetime
\be\label{Flow_u_Thermal_half}
\frac
{
e^{\frac{4\pi}{\beta}x}-2e^{\frac{2\pi}{\beta}(x-L)}\cosh\frac{2\pi}{\beta}t+e^{-\frac{2\pi}{\beta}2L}
}
{
e^{\frac{4\pi}{\beta}x}-2e^{\frac{2\pi}{\beta}(x+L)}\cosh\frac{2\pi}{\beta}t+e^{\frac{2\pi}{\beta}2L}
}
=e^{2u}.
\ee

Shown in Fig.$\,$$\,$\ref{ThermalState_Half} are the constant-$u$ flows for different $\beta$,
plotted according to Eq.\ (\ref{Flow_u_Thermal_half}).
One can find that these constant-$u$ flows are equally distributed vertical lines.
In addition, the density of these flows is proportional to $\beta^{-1}$.
To understand these features, let us focus on the causal wedge of $A$ defined by
\be
x\ge 0, \quad t>x-L,\quad t<-(x-L).
\ee
In the limit $L, x, t\gg \beta$, Eq.\ (\ref{Flow_u_Thermal_half}) can be simplified as
\be\label{Vertical_FiniteT_Half}
e^{\frac{4\pi}{\beta}(x-L)}=e^{2u}, \quad \Rightarrow\quad x=\frac{\beta}{2\pi}u+L,
\ee
which describes the vertical flows in Fig.$\,$$\,$\ref{ThermalState_Half}.
In addition, one has $\Delta x=\frac{\beta}{2\pi}\Delta u$, and $1/\Delta x=\frac{2\pi}{\beta}\cdot \frac{1}{\Delta u}$,
\textit{i.e.}, the density of these vertical lines is proportional to $\beta^{-1}$.

\subsection{A semi-infinite interval $A$ in an infinite system after a global quench}

\begin{figure}
\begin{center}
\includegraphics[width=4.5in]{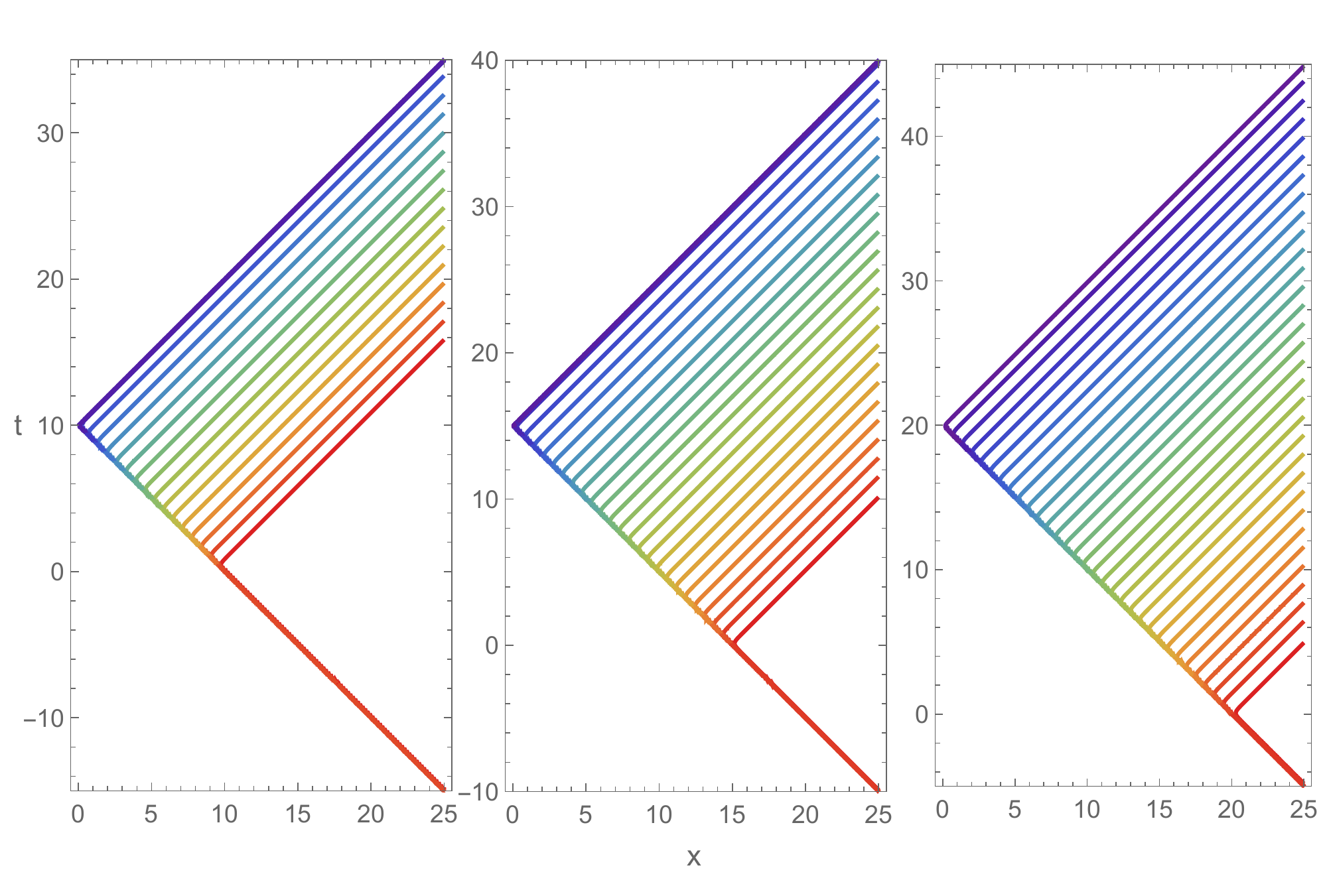}
\end{center}
\caption{
Constant-$u$ flows in the causal wedge of subsystem $A=\{(x,t_0), 0\le x\le \infty\}$ in an infinite system after a global quantum
quench.
The parameters we use are $\beta=1.5$. The observation times are $t_0=10,15$ and $20$ from left to right.
}\label{GlobalFlowInfinite}
\end{figure}

The setup for a global quench in CFT can be described by
the infinite strip given by $-\beta/4\le \text{Im}(z)\le \beta/4$ and $\text{Re}(z)\in \mathbb{R}$.
We are interested in  subsystem $A=(0,\infty)$, and therefore need to consider
a cut  $C=\{z=i\tau+x,x\ge 0\}$, where
$|\tau|<\beta/4$. The conformal transformation is (as compared to Ref.$\,$\cite{Cardy1608},
there is a sign difference here, introduced to simplify  comparison 
 with the conformal mapping used in the main text)
\begin{equation}
w=-\log \frac{\sinh\frac{\pi(z-i\tau)}{\beta}}{\cosh\frac{\pi(z+i\tau)}{\beta}},
\end{equation}
based on which we can find the constant-$u$ flows in Euclidean spacetime
\begin{equation}\label{F_Eu01}
\frac{\sinh^2\frac{\pi x}{\beta}\cos^2\frac{\pi(y-\tau)}{\beta}+\cosh^2\frac{\pi x}{\beta}\sin^2\frac{\pi(y-\tau)}{\beta}}
{\cosh^2\frac{\pi x}{\beta}\cos^2\frac{\pi(y+\tau)}{\beta}+\sinh^2\frac{\pi x}{\beta}\sin^2\frac{\pi(y+\tau)}{\beta}}
=e^{-2u},
\end{equation}
and (by taking $y\to it$ and $\tau\to it_0$)
\begin{equation}\label{F_Eu}
\frac{\sinh^2\frac{\pi x}{\beta}\cosh^2\frac{\pi(t-t_0)}{\beta}-\cosh^2\frac{\pi x}{\beta}\sinh^2\frac{\pi(t-t_0)}{\beta}}
{\cosh^2\frac{\pi x}{\beta}\cosh^2\frac{\pi(t+t_0)}{\beta}-\sinh^2\frac{\pi x}{\beta}\sinh^2\frac{\pi(t+t_0)}{\beta}}
=e^{-2u}
\end{equation}
in Minkowski spacetime.
The constant-$u$ flows corresponding to subsystem $A$ are shown in Fig.\ref{GlobalFlowInfinite}.
As $t_0$ grows, the region $//$ filled with right-tilted lines grows all the way, due to the semi-infinite
nature  of both subsystems $A$ and $B$.

Equation (\ref{F_Eu}) can be further simplified as
\be\label{Global_Homogeneous}
\frac
{
\cosh\frac{2\pi x}{\beta}-\cosh\frac{2\pi(t-t_0)}{\beta}
}
{
\cosh\frac{2\pi x}{\beta}+\cosh\frac{2\pi(t+t_0)}{\beta}
}=e^{-2u}.
\ee
Let us check the region $//$ filled with right-tilted lines, which is defined by
\be
t-t_0<x, \quad t-t_0>-x, \quad t-t_0>x-2t_0.
\ee
In the limit $t\pm t_0\gg \beta$, then Eq.\ (\ref{Global_Homogeneous}) can be approximated as
\be\label{Global_infinity_left_tilted}
\frac
{
e^{\frac{2\pi x}{\beta}}
}
{
e^{\frac{2\pi(t+t_0)}{\beta}}
}=e^{-2u}\quad
\Rightarrow
x-(t+t_0)=-\frac{\beta}{\pi}u.
\ee
Similarly, if we study the flows for subsystem $B=(-\infty,0)$, one can observe a region $\backslash\backslash$
filled with left-tilted flows. This region is defined by
\be
t-t_0>x, \quad t-t_0<-x, \quad t-t_0>-(x+2t_0).
\ee
In the limit $t\pm t_0\gg \beta$, Eq.\ (\ref{Global_Homogeneous}) can be approximated as
\be\label{Left_tilted_B}
\frac
{
e^{\frac{-2\pi x}{\beta}}
}
{
e^{\frac{2\pi(t+t_0)}{\beta}}
}=e^{-2u}\quad
\Rightarrow
x+(t+t_0)=\frac{\beta}{\pi}u.
\ee

\bibliography{QuenchEE}

\section*{References}

\begin{thebibliography}{99}




\bibitem{Page1993}
D.~N.~Page,
Phys. Rev. Lett. 71, 1291 (1993).

\bibitem{Deutsch1991}
J.~M.~Deutsch,
Phys. Rev. A 43, 2046 (1991).

\bibitem{Jensen1985}
 R.~V.~Jensen, R. Shankar, Phys. Rev. Lett. 54, 1879–1882 (1985).

\bibitem{Srednicki1994}
M.~Srednicki,
Phys. Rev. E 50, 888 (1994).

\bibitem{Rigol2008}
 M.~Rigol, V.~Dunjko, M.~Olshanii,
Nature 452, 854 (2008).

\bibitem{Eisert2015}
J.~Eisert, M.~Friesdorf, C.~Gogolin,
Nat. Phys. 11, 124 (2015); arXiv:1408.5148.

\bibitem{Kaufman2016}
A.~M.~Kaufman,\ \textit{et al}.,
Science 353, 794 (2016).





\bibitem{Rigol2007}
 M. Rigol, V. Dunjko, V. Yurovsky, and M. Olshanii,
Phys. Rev. Lett. 98, 050405 (2007); arXiv:cond-mat/0604476.

\bibitem{GGE}
See, e.g.,
M. A. Cazalilla,
Phys. Rev. Lett. 97, 156403 (2006); arXiv:cond-mat/0606236;\\
M. Cramer, C. M. Dawson, J. Eisert, and T. J. Osborne,
Phys. Rev. Lett. 100, 030602 (2008); arXiv:cond-mat/0703314;\\
P.~Calabrese, F.~H.~L.~Essler, and M.~Fagotti,
Phys. Rev. Lett. 106, 227203 (2011); arXiv:1104.0154;\\
J.~-S.~Caux and R.~M.~Konik,
Phys. Rev. Lett. 109, 175301 (2012); arXiv:1203.0901;\\
M.~Fagotti and F.~H.~L.~Essler,
Phys. Rev. B 87, 245107 (2013); arXiv:1302.6944\\
J.-S.~Caux and F.~H.~L.~Essler,
Phys. Rev. Lett. 110, 257203 (2013); arXiv:1301.3806\\
M.~Kormos, M.~Collura, and P.~Calabrese,
Phys. Rev. A 89, 013609 (2014); arXiv:1307.2142.



\bibitem{RyuHatsugai2006}
  S.\ Ryu and Y.\ Hatsugai,
  Phys. Rev. B 73, 245115 (2006); arXiv:cond-mat/0601237.

\bibitem{Haldane2008}
H. Li and F.~D.~M. Haldane,
Phys. Rev. Lett. 101, 010504 (2008); arXiv:0805.0332.

\bibitem{Top_1D_0910}
F.~Pollmann, E.~Berg, A.~M.~Turner and Masaki Oshikawa,
Phys. Rev. B 81, 064439 (2010); arXiv:0910.1811.

\bibitem{Katsura}
X.-L. Qi, H. Katsura, A. W. W. Ludwig,
Phys. Rev. Lett. 108 (2012) 196402.

\bibitem{BauerVidal}
B. Bauer, L. Cincio, B. P. Keller, M. Dolfi, G. Vidal, S. Trebst, A. W. W. Ludwig,
Nature Communications 5 (2014) 5137.



\bibitem{blanco2013}
D. Blanco, H. Casini, L.-Y. Hung and R. Myers,
JHEP 1308 (2013) 060; arXiv:1305.3182.

\bibitem{peschel-04} I. Peschel,
J. Stat. Mech. P12005 (2004).

\bibitem{bw}
J. Bisognano and E. Wichmann,
J. Math. Phys. 17, 303 (1976);
J. Math. Phys. 16, 985 (1975).

\bibitem{unruh}
W. Unruh,
Phys. Rev. D 14, 870 (1976).



\bibitem{chm}
H. Casini, M. Huerta and R. Myers,
JHEP 1105 (2011) 036; arXiv:1102.0440;\\
P. D. Hislop and R. Longo, 
Commun. Math. Phys. 84, 71 (1982).



\bibitem{Wong2013}
G.~Wong, I.~Klich, L.~Pando Zayas and D.~Vaman, JHEP 1312 (2013) 020; arXiv:1305.3291.

\bibitem{Casini0903}
H. Casini and M. Huerta,
Class. Quant. Grav. 26, 185005 (2009); arXiv:0903.5284.

\bibitem{Longo0912}
R. Longo, P. Martinetti and K. H. Rehren,
Rev. Math. Phys. 22, 331 (2010); arXiv:0912.1106.

\bibitem{Wong1805}
G. Wong, arXiv:1805.10651.

\bibitem{Peschel1999_2009}
I.\,Peschel, M.\,Kaulke, and O.\, Legeza, Ann. der Phys. 8, 153 (1999);\\
I. Peschel and V. Eisler, J. Phys. A: Math. Theor. 42, 504003 (2009); arXiv:0906.1663.


\bibitem{Eisler1703}
V.\,Eisler, I.\,Peschel, 
J. Phys. A: Math. Theor. 50 284003 (2017); 
arXiv:1703.08126;\\
V.\,Eisler, I.\,Peschel, 
arXiv:1805.00078;

\bibitem{ToldinAssaad1804}
F.\,P.\, Toldin, F.\,F.\,Assaad, arXiv:1804.03163.

\bibitem{He1805}
W.\,Zhu, Z.\,Huang, Y.\,-C.\,He, arXiv:1806.08060.


\bibitem{EH1807}
G.\,Giudici, T.\,Mendes-Santos, P.\,Calabrese, M.\,Dalmonte, arXiv:1807.01322.



\bibitem{Cardy1608}
J. Cardy and  E. Tonni,
 J. Stat. Mech. 123103 (2016); arXiv:1608.01283.

\bibitem{HartmanJHEP2015}
C. T. Asplund, A. Bernamonti, F. Galli, T. Hartman,
JHEP09 (2015) 110; arXiv:1506.03772

\bibitem{Cardy2005}
P.~Calabrese and J.~Cardy,
J. Stat. Mech. P04010 (2005); arXiv:cond-mat/0503393.

\bibitem{Cardy1603}
P.\,Calabrese, J.\,Cardy,
J. Stat. Mech. (2016) 064003; arXiv:1603.08267.



\bibitem{Calabrese2006}
P.\,Calabrese and J.\,Cardy,
 Phys. Rev. Lett. 96 136801 (2006); arXiv:cond-mat/0601225\\
P.\,Calabrese and J.\,Cardy,
J. Stat. Mech. 0706 P008 (2007); arXiv:0704.1880.



\bibitem{Cardy2015}
J.\, Cardy,
J. Stat. Mech. (2016) 023103; arXiv:1507.07266

\bibitem{Mandal2015}
G.\,Mandal, R.\,Sinha and N.\,Sorokhaibam,
JHEP 08 (2015) 013; arXiv:1501.04580.


\bibitem{Miyaji}
M, Miyaji, S. Ryu, T. Takayanagi and X. Wen,
JHEP 05 (2015) 152; arXiv:1412.6226.




\bibitem{Cardy1507}
J.~Cardy,
 J. Stat. Mech. 023103 (2016); arXiv:1507.07266.

\bibitem{Affleck-Ludwig}
I.\,Affleck and A.\,Ludwig,
Phys. Rev. Lett. 67 (1991) 161.

\bibitem{Cardy1403}
J.$\,$Cardy,
Phys. Rev. Lett. 112, 220401 (2014); arXiv:1403.3040.


\bibitem{Mandal1604}
K.$\,$Kuns and  D.$\,$Marolf,
JHEP 09 (2014) 082; arXiv:1406.4926;\\
G. Mandal, R. Sinha, T. Ugajin,
arXiv:1604.07830.

\bibitem{ModularFlowThermal}
H.~J.~Borchers and J.~Yngvason
Journal of Mathematical Physics 40, 601 (1999).

\bibitem{CardyVerlindeNPB1989}
John~L.Cardy
Nucl. Phys. B324 (1989) 581.


\bibitem{InhomogeneousQuench}
See, e.g.,
S.\,Sotiriadis and J.\,Cardy,  J. Stat. Mech P11003 (2008); arXiv:0808.0116 \\
D.\,Bernard and B.\,Doyon J. Phys. A 45 362001 (2012); arXiv:1202.0239\\
J.\,Viti, J.-M.\,Stephan, J.\,Dubail and M.\,Haque, EPL 115 (2016) 40011; arXiv:1507.08132;\\
J.\,Dubail, J.\,Stephan,\,J.\,Viti and P.\,Calabrese, SciPost Phys. 2, 002 (2017); arXiv:1606.04401;\\
K.\,Agarwal, E.\,G.\,D.\,Torre, J.\,Schmiedmayer and E.\,Demler, Phys. Rev. B 95, 195157 (2017); arXiv:1609.04046;\\
G.\,Ramírez, J.\,Rodríguez-Laguna, and G.\,Sierra, J. Stat. Mech. (2015) P06002; arXiv:1503.02695.\\
J.\,Rodríguez-Laguna, J.\,Dubail, G.\,Ramírez, P.\,Calabrese, and G.\,Sierra, 
 J. Phys. A: Math. Theor. 50 164001 (2017); arXiv:1611.08559.\\
J.\,Dubail, J.-M,\,Stephan, and P.\,Calabrese, 	SciPost Phys. 3, 019 (2017); arXiv:1705.00679;\\
V.\,Eisler, D.\, Bauernfeind, Phys. Rev. B 96, 174301 (2017); arXiv:1708.05187;\\
X.\, Wen, Y.\, Wang, and S.\,Ryu, J. Phys. A: Math. Theor. 51 195004 (2018); arXiv:1711.02126;\\
X.\, Wen, J.\,-Q.\,Wu, Phys. Rev. B 97, 184309 (2018); arXiv:1802.07765.

\bibitem{Tonni2017}
E.~Tonni, J.~Rodriguez-Laguna, and G.~Sierra, 
J. Phys. A: Math. Theor. 4 043105 (2018);
arXiv:1712.03557.

\end{thebibliography}

\end{appendices}

\end{document}